\documentclass[structabstract]{aa}  
\usepackage{graphicx}
\usepackage[round]{natbib}
\usepackage{url}
\bibpunct{(}{)}{;}{a}{}{,} 
\usepackage{txfonts}

\def\hc3n{HC$_3$N}
\def\h3op{H$_3$O$^+$}

\def\cm3{cm$^{-3}$}
\def\cm2{cm$^{-2}$}

\def\s1{s$^{-1}$}

\def\asec{$''$}

\begin{document}
   \title{Molecules as tracers of galaxy evolution: an EMIR survey}

   \subtitle{I. Presentation of the data and first results}

   \author{F.~Costagliola
          \inst{1}\fnmsep\thanks{F.C. wishes to thank the EU ESTRELA programme for support.}
          \and
          S.~Aalto\inst{1}\fnmsep\thanks{S.A. wishes to thank the Swedish Research Council for grant support.}
	\and
	M.~I.~Rodriguez\inst{2}
	\and
	S.~Muller\inst{1}	
	\and
	H.~W.~W.~Spoon\inst{3}
	\and
	S.~Mart{\'i}n\inst{4}
 	\and
	M.~A.~Per{\'e}z-Torres\inst{2}
	\and
	A.~Alberdi\inst{2}
	\and
	J.~E.~Lindberg\inst{1,5}
	\and
	F.~Batejat\inst{1}
	\and
	E.~J{\"u}tte\inst{6}
	\and
	P.~van~der~Werf\inst{7,8}
	\and
	F.~Lahuis\inst{7,9}
	}
	\institute{Department of Earth and Space Sciences, Chalmers University of Technology, Onsala Space Observatory, SE-439 92 Onsala, Sweden, 
        \email{francesco.costagliola@chalmers.se}
        \and Instituto de Astrof{\'i}sica de Andaluc{\'i}a (IAA-CSIC), PO Box 3004, E-18080 Granada,  Spain
        \and Cornell University, Astronomy Department, Ithaca, NY 14853, USA
	\and European Southern Observarory, Alonso de C{\'o}rdova 3107, Vitacura, Casilla 19001, Santiago 19, Chile
	\and Centre for Star and Planet Formation, Natural History Museum of Denmark, University of Copenhagen, {\O}ster Voldgade 5-7, 1350 K{\o}benhavn K, Denmark
	\and Astronomisches Institut Ruhr-Universitaet Bochum, Universitaetsstr. 150, 44780 Bochum, Germany
	\and Leiden Observatory, Leiden University, NL-2300 RA, Leiden, The Netherlands   
	\and Institute for Astronomy, University of Edinburgh, Royal Observatory, Blackford Hill, Edinburgh EH9 3HJ, United Kingdom
	\and SRON Netherlands Institute for Space Research, P.O. Box 800, NL-9700 AV, Groningen, The Netherlands               
}

   \date{}

 
  \abstract
   {}
   {We investigate the molecular gas properties of a sample of 23 galaxies in order to find and test
chemical signatures of galaxy evolution and to compare them to IR evolutionary tracers.}
   { Observation at 3 mm wavelengths were obtained with the EMIR broadband receiver, mounted on the IRAM 30~m telescope on Pico Veleta, Spain. We compare the emission of the main molecular species with existing models of chemical evolution by means of line intensity ratios diagrams and principal component analysis. }
   {We detect molecular emission in 19 galaxies in two 8 GHz-wide bands centred at 88 and 112 GHz. The main detected transitions are the $J$=1--0 lines of CO, $^{13}$CO, HCN, HNC, HCO$^+$, CN, and C$_2$H. We also detect HC$_3$N $J$=10--9 in the galaxies IRAS~17208, IC~860, NGC~4418, NGC~7771, and NGC~1068. The only HC$_3$N detections are in objects with HCO$^+$/HCN$<$1 and warm IRAS colours. Galaxies with the highest HC$_3$N/HCN ratios have warm IRAS colours (60/100 $\mu$m$>$0.8). The brightest HC$_3$N emission is found in IC~860, where we also detect the molecule in its vibrationally excited state. We find low HNC/HCN line ratios ($<$0.5), that cannot be explained by existing PDR or XDR chemical models. The intensities of HCO+ and HNC appear anti-correlated, because galaxies with low HCO+/HCN intensity ratios have high HNC/HCN. No correlation is found between the HNC/HCN line ratio and dust temperature. All HNC-bright objects are either luminous IR galaxies (LIRG) or Seyferts. Galaxies with bright polycyclic aromatic hydrocarbons (PAH) emission show low HNC/HCO$^+$ ratios.  The CO/$^{13}$CO ratio is positively correlated with the dust temperature and is generally higher than in our galaxy. The emission of CN and C$^{18}$O is correlated.}
   {Bright HC$_3$N emission in HCO$^+$-faint objects may imply that these are not dominated by X-ray chemistry. Thus the HCN/HCO$^+$ line ratio is not, by itself, a reliable tracer of XDRs. Bright HC$_3$N and faint HCO$^+$ could be signatures of embedded star-formation, instead of AGN activity. Mechanical heating caused by supernova explosions may be responsible for the low HNC/HCN  and high HCO$^+$/HCN ratios in some starbursts. We cannot exclude, however, that the discussed trends are largely caused by optical depth effects or excitation. Chemical models alone cannot explain all properties of the observed molecular emission. Better constraints to the gas spacial distribution and excitation are needed to distinguish abundance and excitation effects.}

\titlerunning{EMIR Molecular Tracers}
   \maketitle
%

\section{Introduction}

Luminous infrared galaxies (LIRGs) radiate most of their luminosity (L$_\mathrm{IR}>$10$^{11}$~L$_{\odot}$) as
dust thermal emission in the infrared and have been studied at almost all wavelenghts \citep{sanders_96}. 
However, the nature of the power source is still unclear when the inner region of the LIRG is obscured
by dust. The high central IR surface brightness implies that this power source can be either an embedded compact starburst
or an enshrouded AGN - or a combination of both.
The evolution of the activity and the connection between AGN and starburst are still not  well understood
and must be further explored. At high  redshift, LIRGs dominate the cosmic infrared background and, by
assuming that they are powered by starburts, we can use these galaxies to trace the dust-obscured star-formation rate, the dust content and the metallicity in the early Universe \citep{barger99}. The most well known
techniques to distinguish between AGN and star-powered galaxies rely on the observation of emission lines
in the optical \citep[e.g., ][]{vo87}. Over the last decade several other diagnostic diagrams, based on
IR spectra, have been proposed to quantify the contribution of star-formation and AGN activity to the
infrared luminosities of LIRGs \citep{genzel98,lutz98,spoon07}. 
In the mm and sub-mm there have been attempts as well to classify the activity of galaxies via diagnostic
diagrams, of which the most well known is the HCN/HCO$^+$ line ratio plot of \citet{kohno2001} and
\citet{imanishi04}.  A multi-transition study of the HCN/HCO$^+$ ratio in Seyfert and starburst galaxies has
also been reported by \citet{krips08}. These authors find an underluminosity of HCO$^+$ in some AGN-dominated
cores, which they suggested to be owing to an underabundance of HCO$^+$ caused by the X-ray-dominated chemistry
induced by the AGN \citep{maloney96}. This latter interpretation has been disputed since it does not agree
with recent models of X-ray-dominated regions (XDRs) \citep{meijerink05}, which  instead show an
enhancement of HCO$^+$ abundances caused by an increase
in ionization. Other interpretations in terms of starburst evolution have been put forward as an alternative
\citep[e.g., ][]{baan08}. \citet{gracia06} also found that ULIRGs (L$_\mathrm{IR}>$10$^{12}$~L$_{\odot}$) in general seem to have lower HCO$^+$ 1--0
luminosities with regards to HCN - compared with more moderate LIRGs. The HCN/CN 1--0 and HCN/HNC 1--0 line ratios have been used as well to help interpret
a galaxy's position in an evolutionary scheme \citep{aalto02,baan08}. For the brightest nearby galaxies, a
scheme based on the HNCO/CS ratio has been proposed \citep{martin09}. Moreover, prominent sources such as
NGC~253 and IC~342 allow the detection of rarer species that can be used in turn to help identifying and even resolving
the dominant activity  \citep[e.g., ][]{meier05,martin06}.\\
Line ratios need to be very accurately measured if we want them to be sensitive tracers of molecular properties. The new EMIR receiver, mounted in 2009 on the IRAM 30 m telescope in Spain, offers the opportunity of achieving this high accuracy. The available bandwidth of nearly 8 GHz at 3 mm allows us to fit  many key molecular lines in the same band, therefore eliminating the uncertainties owing to relative calibration and pointing errors that affect most single-dish observations.
We present a survey of molecular lines for a sample of 23  galaxies observed with EMIR in the period June-November 2009. In section \ref{sec:obs} we report the details about the observations and source selection. In Sect. \ref{sec:results} we present the results. In Sect. \ref{sec:discussion} the line ratio diagrams for HCN, HNC, HCO$^+$, and HC$_3$N are presented and discussed. The molecular emission is also compared with far-infrared (FIR) colours and polycyclic aromatic hydrocarbons (PAH) emission. In Sect. \ref{sec:conclusions} we present our conclusions and an outlook. The spectra of all observed sources and the tables summarizing the line parameters of the detected species can be found in the Appendix.
 
\begin{figure}[h]
\label{fig:spoondiag}
\centering
\includegraphics[width=.5\textwidth,keepaspectratio]{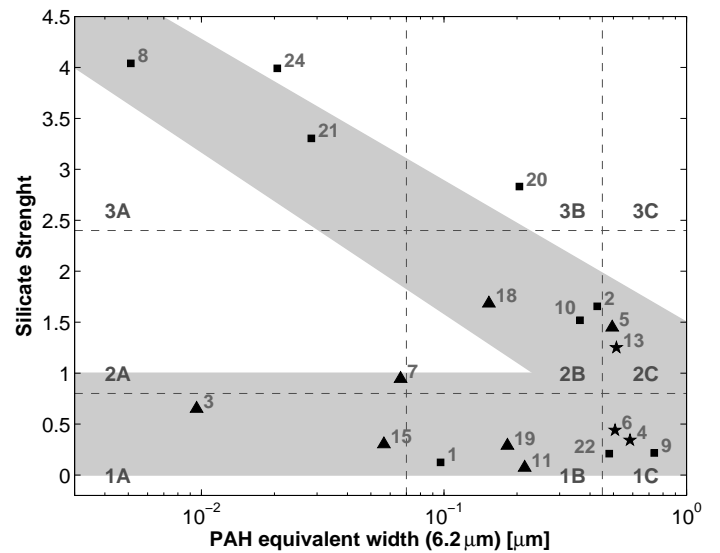}
\caption{Position of the observed galaxies on the mid-infrared diagnostic diagram from \citet{spoon07}. Galaxies are divided into nine classes ({\it 1A--3C}), depending on their spectral properties and position on the graph. These classes range from PAH-dominated spectra ($1C$) to continuum-dominated spectra with faint PAH emission ($1A$) and obscured galaxies with deep silicate absorption ($3A$). Galaxy types are distinguished by their plotting symbol: {\it Squares:} LIRGs. {\it Stars:} Starburst galaxies. {\it Triangles:} Seyfert galaxies. The numbers indicate different galaxies, as explained in Table \ref{tab:linrat}. The shadowed area corresponds to the two main sequences discussed by \citet{spoon07}. An increase of the PAH equivalent width indicate an increase of the starburst contribution to the emission, while AGN spectra are continuum-dominated. See Section \ref{sec:sel} for further discussion.
}
\end{figure}

\begin{table}[h]
\renewcommand{\arraystretch}{1.2}
\setlength{\tabcolsep}{5pt}
{\tiny
\label{tab:bands}
\centering
\begin{tabular}{l c c c c c} 
\hline 
\hline
\\ 
Line & HPBW & $\eta_\mathrm{mb}$ & $\nu_0$ & E$_\mathrm{up}$ & n$_\mathrm{c}$ (20 K)\\
 & [ \asec ] & & [ GHz ] & [ K ] & [ cm$^{-3}$ ] \\
\\
\hline
\\
88 GHz Band & 29 & 0.81 & & & \\
\\
SiO (2-1) & $''$ & $''$ & 86.847 & 6.25 & 1$\times$10$^5$\\
C$_2$H (1-0) & $''$ & $''$&  87.316 & 4.19 & 1$\times$10$^5$\\
HCN (1-0) &$''$ &$''$ &  88.633 & 4.25 & 2$\times$10$^5$\\
HCO$^+$ (1-0) &$''$ &$''$ &  89.188 & 4.28 & 3$\times$10$^4$\\
HNC (1-0) &$''$ &$''$ &  90.663 & 4.35 & 1$\times$10$^5$\\
HC$_3$N (10-9) & $''$&$''$ &  90.978 & 24.0 & 9$\times$10$^4$\\
\\
112 GHz Band & 22 &  0.78 & & &  \\
\\
HC$_3$N (12-11) & $''$ & $''$& 109.173 & 34.0 & 2$\times$10$^5$\\
C$^{18}$O (1-0) &$''$ &$''$ & 109.757 & 5.27 &  4$\times$10$^2$\\
$^{13}$CO (1-0) &$''$ &$''$ &  110.201 & 5.29 &  4$\times$10$^2$\\
CN (1-0)J=1/2-1/2 &$''$  &$''$ & 113.191 & 5.43 & 2$\times$10$^6$\\
CO (1-0) & $''$ & $''$& 115.271 & 5.53 & 4$\times$10$^2$\\
\\
\hline
\end{tabular}}

\caption{\label{tab:bands}Properties of the two observed bands. Rest frequencies of the brightest lines in each band are shown with the beam sizes (HPBW) and  main beam efficiencies ($\eta_\mathrm{mb}$). Energies of upper level (E$_\mathrm{up}$) and critical densities at 20 K ($n_\mathrm{c}$) were taken from the {\it Leiden atomic and molecular database}.}
\end{table}

\section{Source selection}
\label{sec:sel}
The first diagnostic diagram to take into account the effects of strong obscuration of the nuclear power source was
presented by \citet{spoon07}, using the equivalent width of the 6.2 $\mu$m PAH emission feature and the strength of
the 9.7  $\mu$m silicate absorption (see Fig. \ref{fig:spoondiag}). \\
Based on the position in the diagram, galaxies are put into nine classes, ranging from continuum-dominated AGN hot
dust spectra ($1A$) to PAH-dominated starburst spectra ($1C$) to absorption-dominated spectra of deeply obscured galactic
nuclei ($3A$). \\
\citet{spoon07} find that galaxies are systematically distributed along two distinct branches: one horizontal sequence
at low silicate  depths, ranging from AGN to starburst-dominated spectra, and one diagonal sequence at higher silicate
strength, ranging from obscured nuclei to pure-starburst objects. The separation into two branches likely reflects
fundamental differences in the dust geometry in the two sets of sources. Spectra of luminous infrared galaxies are found
along the full length of both branches, reflecting the diverse nature of the LIRG family. In this work, the term {\it LIRG} broadly refers to objects which emit most of their energy in the IR. Thus a galaxy classified as LIRG may be a starburst, an AGN, or both, but the
dust obscuration hinders a clear classification. We do not explicitly distinguish between LIRGs (L$_\mathrm{IR}>$10$^{11}$~L$_{\odot}$) and ULIRGs (L$_\mathrm{IR}>$10$^{12}$~L$_{\odot}$). However, IR luminosities are reported in Table \ref{tab:linrat} for reference. Objects with both LIRG and AGN signatures are labelled as $A,L$ in Table \ref{tab:linrat} and appear as AGNs ($triangles$) in the graphs. \\
\citet{spoon07} interpret the observed distribution as a possible evolutionary effect, with sources moving from the
diagonal to the horizontal branch as the dust distribution evolves from a uniform to a clumpy geometry. The underpopulated
2A class implies either that this transition for LIRGs is very rapid, or that LIRGs mostly evolve into unobscured starbursts.\\
We aim to compare this mid-IR evolution scheme with mm molecular observations. Our targets were mainly
selected from the sample of \citet{spoon07}, plus a few interesting objects for which we had PAH and silicate mid-IR data.
The source selection criteria were the following:    
{\it
\begin{itemize}
\item Uniform coverage of the most significant classes in the diagnostic diagram of \citet{spoon07}, 
\item Uniform representation of different galaxy types (Seyferts, starbursts, LIRGs),
\item Source visibility at IRAM site.
\end{itemize}
}
The resulting sample is composed of 23 galaxies, whose main properties are listed in Tab \ref{tab:linrat}. All targets were observed in the 88 GHz band, but because of time constraints, only 12 were observed in the 112 GHz band.

\begin{table*}
\centering
\renewcommand{\arraystretch}{1.2}
\setlength{\tabcolsep}{5pt}
{\tiny
\begin{tabular}{lccccccccclll} 
\hline 
\hline 
\\ 
Galaxy & RA & Dec & V$_\mathrm{Helio}$ & ${\rm \frac{HCO^+(1-0)}{HCN(1-0)}}$ & ${\rm\frac{HNC(1-0)}{HCN(1-0)}}$ & ${\rm\frac{HC_3N(10-9)}{HCN(1-0)}}$ & ${\rm\frac{C_2H(1-0)}{HCN(1-0)}}$ & ${\rm\frac{CO(1-0)}{^{13}CO(1-0)}}$ & Log(L$_{\rm IR}$/L$_\odot$) & Class & Type & Num. \\ 
 & $[h:m:s]$ & $[^{\circ}:':'']$ & $[km/s]$ & & & & & & & & &\\[5pt] 

\hline \\ 
IRAS~17208 & 17:23:21.9 & -00:17:01 & 12834 & 0.78(0.12)  & 0.77(0.14)  & 0.31(0.07)  & 0.42(0.10)  & -  & 12.35 & 1B & L & 1 \\ 
 IC~860 & 13:15:03.5 & +24:37:08 & 3347 & 0.62(0.16)  & 0.62(0.16)  & 0.42(0.13)  & 1.04(0.29)  & 18.96(3.83)  & 11.14 & 2B & L & 2 \\ 
 Mrk~231 & 12:56:14.2 & +56:52:25 & 12642 & 0.56(0.08)  & 0.38(0.06)  & $<$0.07  & 0.29(0.12)  & -  & 12.37  & 1A & A,L & 3 \\ 
 NGC~1614 & 04:33:59.8 & -08:34:44 & 4778 & 1.83(0.37)  & 0.33(0.15)  & $<$0.35  & 1.00(0.33)  & 28.81(2.10) & 11.43 & 1C & S & 4 \\ 
 NGC~3079 & 10:01:57.8 & +55:40:47 & 1116 & 1.12(0.11)  & 0.27(0.05)  & $<$0.06  & 0.54(0.09)  & 17.10(1.16) & 10.65 & 2C & A & 5 \\ 
 NGC~4194 & 12:14:09.5 & +54:31:37 & 2501 & 1.32(0.35)  & 0.53(0.23)  & $<$0.20  & 0.98(0.36)  & 19.10(1.90) & 10.93  & 1C & S & 6 \\ 
 NGC~4388 & 12:25:46.7 & +12:39:44 & 2524 & 1.38(0.40)  & 0.62(0.24)  & $<$0.35  & 0.81(0.52)  & -  & 9.66 & 2A & A & 7 \\ 
 NGC~4418 & 12:26:54.6 & -00:52:39 & 2110 & 0.59(0.10)  & 0.47(0.09)  & 0.37(0.09)  & 0.64(0.18)  & - & 11.00  & 3A & L & 8 \\ 
 NGC~6090 & 16:11:40.7 & +52:27:24 & 8785 & 1.67(0.44)  & 0.25(0.17)  & $<$0.18  & 0.46(0.20)  & - & 11.34  & 1C & L & 9 \\ 
 NGC~6240 & 16:52:58.9 & +02:24:03 & 7339 & 1.63(0.14)  & 0.20(0.06)  & $<$0.09  & 0.36(0.08)  &  28.79(2.99)  & 11.69 & 2B & L & 10 \\ 
 NGC~7469 & 23:03:15.6 & +08:52:26 & 4892 & 1.12(0.11)  & 0.55(0.08)  & $<$0.07  & 0.75(0.24)  & 20.80(0.47) & 11.41  & 1B & A & 11 \\ 
 NGC~7771 & 23:51:24.9 & +20:06:43 & 4277 & 0.94(0.09)  & 0.44(0.07)  & 0.06(0.04)  & 0.32(0.06)  & 13.61(0.52)  & 11.24 & - & L & 12 \\ 
 NGC~660 & 01:43:02.4 & +13:38:42 & 850 & 1.04(0.09)  & 0.52(0.06)  & $<$0.07  & 0.45(0.08)  & 16.45(0.49)  & 10.40  & 2C & S & 13 \\ 
 NGC~3556 & 11:11:31.0 & +55:40:27 & 699 & 1.57(0.37)  & 0.37(0.24)  & $<$0.27  & 1.56(0.37)  & 12.50(0.31) & 10.00  & - & S & 14 \\ 
 NGC~1068 & 02:42:40.7 & -00:00:48 & 1137 & 0.67(0.03)  & 0.40(0.03)  & 0.04(0.02)  & 0.37(0.09)  & - & 10.89  & 1A & A & 15 \\ 
 NGC~7674 & 23:27:56.7 & +08:46:45 & 8671 & -   & -   & -   & -   & 14.46(1.18) & 11.50 & - & A & 16 \\ 
 UGC~2866 & 03:50:14.9 & +70:05:41 & 1232 & 1.46(0.20) & 0.51(0.13)  & $<$0.14  & 0.88(0.17)  & 20.68(0.80) & 10.69  & - & S & 17 \\ 
 UGC~5101 & 09:35:51.6 & +61:21:11 & 11802 & 0.36(0.20)  & 0.82(0.28)  & $<$0.28  & 0.89(0.37)  & $>$8.63  & 11.87 & 2B & A,L & 18 \\ 
 NGC~2273 & 06:50:08.6 & +60:50:45 & 1840 & 1.05(0.37)  & 1.09(0.38)  & $<$0.89  & $<$0.89  & -  & 10.11 & 1B & A & 19 \\ 
 Arp~220 & 15:34:57.2 & +23:30:09 & 5382 & 0.47(0.07)  & 0.49(0.12)  & 0.19(0.06)  & -   & -  & 12.15 & 3B & L & 20 \\ 
IRAS~15250 & 15:26:59.4 & +35:58:38 & 16535 & - & - & -& - & - & 12.02 & 3A & L & 21 \\
NGC~1140 & 02:54:33.6 & -10:01:40 & 1501 & - & - & -& - & - & 9.50 & 1C & S & 22 \\
NGC~1056 & 02:42:48.3 & +28:34:27 & 1545 & - & - & -& - & -& 9.50 & - & S & 23 \\
NGC~1377 & 03:36:39.1 & -20:54:07 & 1792 &  - & - & -& - & - & 9.63 & 3A & O & 24 \\
\\
 \hline \\ 
\end{tabular}}

\caption{\label{tab:linrat}Line ratios and general properties of the observed galaxies. The {\it mid-IR Class} column refers to the classification by \citet{spoon07}, based on the PAH equivalent width and silicate absorption in the mid-infrared. The column {\it Type} summarizes the properties of the object, distinguishing between starbursts ($S$), Seyferts ($A$), and LIRGs ($L$). The galaxy NGC~1377 is classified as $obscured$ ($O$), because it emits mostly in the IR, but its luminosity is not high enough to be classified as LIRG. For a discussion about source types see Section \ref{sec:sel}. In Figs. \ref{fig:spoondiag}, \ref{fig:plots} and \ref{fig:ratplot112}, the sources are labelled with the numbers reported in column {\it Num}.}
\end{table*}

\begin{figure*}
\centering
\includegraphics[width=.9\textwidth,keepaspectratio]{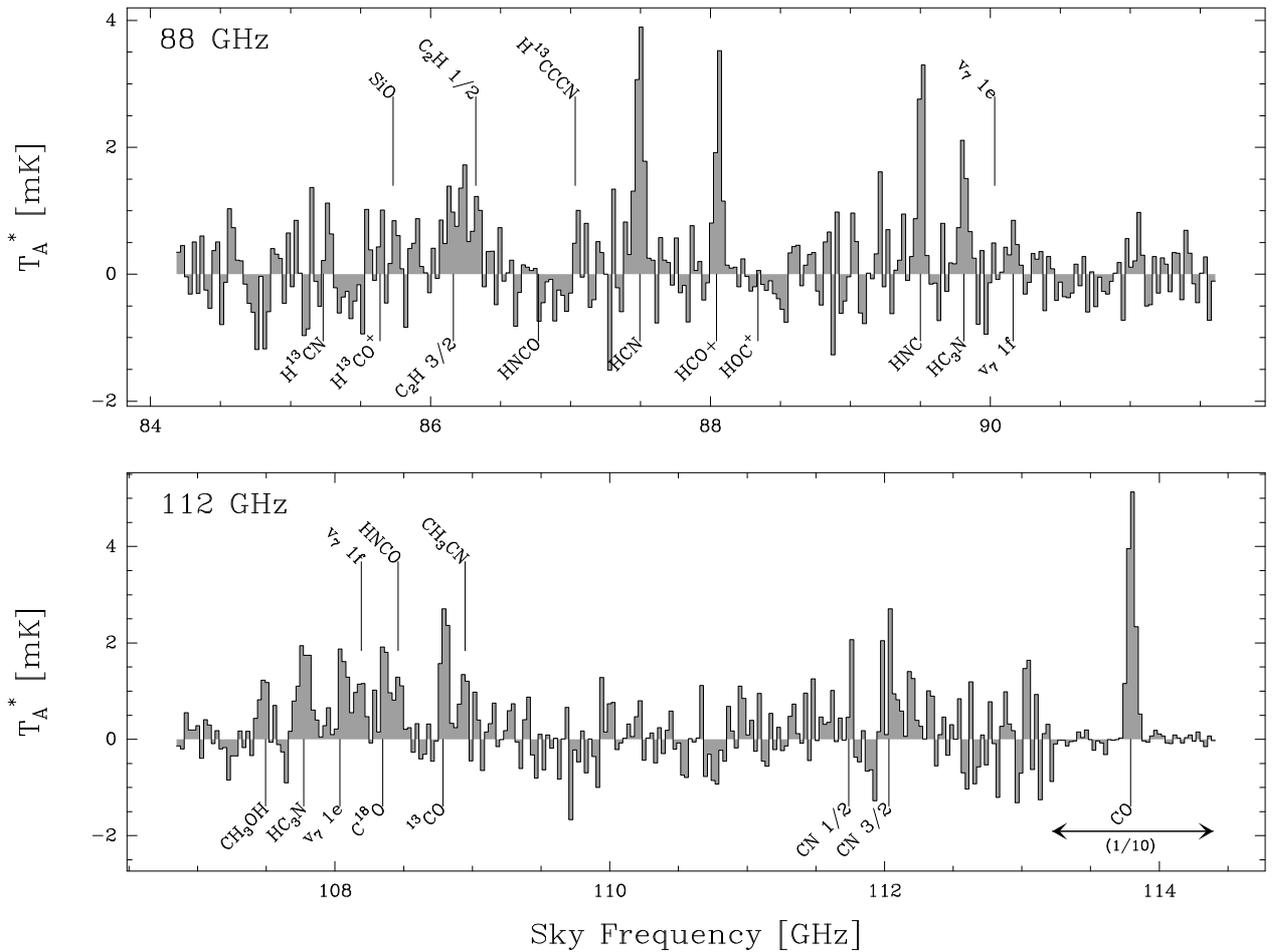}
\caption{\label{fig:ic860}Spectra observed with EMIR in IC~860 at 88 and 112 GHz. The intensity scale is in T$_\mathrm{A}^\star$, not corrected for main beam efficiency. The region marked with $(1/10)$, around the CO 1--0 line, has been scaled down by a factor 10. The main molecular  transitions are labelled regardless of line detection. The C$_2$H 3/2 and 1/2 labels mark the limits of the C$_2$H multiplet at 87 GHz. Transitions of vibrationally excited HC$_3$N are labelled as $v_71e$ and $v_71f$. The frequency scale is the observed frequency, not corrected for redshift.}
\end{figure*}

\section{Observations}
\label{sec:obs}
\subsection{Line ratios at 88 GHz}
\label{sec:88}
The observations were obtained in June-November 2009 with the IRAM 30 m telescope on Pico Veleta, Spain. The 8 GHz band of the EMIR receiver was centred on two different tunings, at 88.675 and 112.15 GHz. These frequencies were chosen in order to accommodate as many potentially strong lines as possible in the same band. A list of the transitions of the most important molecular gas tracers is given in Table \ref{tab:bands}, together with beam efficiencies ($\eta_\mathrm{mb}$) and beam sizes (HPBW) for the two observed bands. Observations were performed in dual polarization, each covering 8 GHz, which nicely fitted the 4$\times$4 GHz backend bottleneck. The E090 frontend was connected to the low-resolution WILMA autocorrelator, which is capable of processing the large input bandwidth with a resolution of about 2 MHz ($\simeq$7 and 5 km s$^{-1}$ at 88 GHz and 112 GHz, respectively). Observations were performed in wobbler switching mode, with a throw of 60-120\asec (depending on source size), in order to maximize baseline quality.\\ The pointing model was checked against bright, nearby calibrators for every source, and every two hours for long integrations. Calibration scans on the standard two load system were taken every 5 minutes. The observing conditions were optimal, with characteristic system temperatures of 100 and 150 K for the 88 GHz and 112 GHz observations, respectively. This resulted in a $rms$ noise per channel of roughly 0.6 mK at 60 km s$^{-1}$ resolution, for 2 hours of on--source observing time. \\
Data were reduced with the CLASS\footnote{http://iram.fr/IRAMFR/GILDAS/} software. A first order baseline was removed from all spectra, which are shown in Appendix \ref{sec:spectra}. The intensity scale in all spectra is in T$_\mathrm{A}^*$, which is related to the source brightness temperature as 
\begin{equation}
T_\mathrm{b}=\frac{T_\mathrm{A}^*}{\eta_\mathrm{mb}}\times\frac{\theta_\mathrm{s}^2+\theta_\mathrm{b}^2}{\theta_\mathrm{s}^2},  
\end{equation}
with $\theta_\mathrm{s}$ and $\theta_\mathrm{b}$ the angular sizes of source and beam, respectively.
We assume a constant main beam efficiency across both observed bands. The error on the main beam brightness temperature estimate introduced by this assumption is of the order of one percent.

\section{Results}
\label{sec:results}

Rest frequencies were taken from the NIST database {\it Recommended Rest Frequencies for Observed Interstellar Molecular Microwave Transitions}\footnote{http://physics.nist.gov/PhysRefData/Micro/Html/contents.html}. Our line identification takes into account the distortion of the velocity scale caused by the large observed bandwidth. For a discussion of this effect, see \citet{gordon92}.\\ 
An example of an EMIR spectrum is shown in Fig. \ref{fig:ic860}, while all the observed spectra are shown in Appendix \ref{sec:spectra}. Line intensities were extracted by means of Gaussian fitting and are reported along with other line parameters in Appendix \ref{sec:tables}.\\
Integrated intensities of the most relevant species were combined to form line intensity ratios, which are listed in Table \ref{tab:linrat}. These were used to produce the diagrams shown in Figs. \ref{fig:plots} and \ref{fig:ratplot112}. We choose to compare intensities of transitions inside the same frequency band to maximize the accuracy of the derived line ratios. The diagrams derived from the two observed bands are described in sections \ref{sec:88} and \ref{sec:112}. A principal component analysis of the 88 GHz dataset is discussed in section \ref{sec:pca}.\\
In plots $a$, $b$, and $c$ of Fig. \ref{fig:plots}, we report the line ratios between the first transitions ($J$=1--0) of HCN, HNC and HCO$^+$ for all detected galaxies. The different symbols refer to different galaxy types.
We note that all non-compact starburst galaxies ({\it stars}) are HCO$^+$-luminous, with line ratios HCO$^+$/HCN $\geq$ 1 (plots $a$, $b$). Luminous infrared galaxies ({\it squares}) and Seyferts ({\it triangles}) do not show any strong trend. However, the majority of active galactic nuclei do reside at low HCO$^+$/HCN values compared with starbursts. \\
The HC$_3$N/HCN line ratio is also reported in the graphs. Circles are drawn around sources where HC$_3$N $J$=10--9 has been detected, the diameter of the circle being proportional to the HC$_3$N/HCN ratio. {\it Evidently that all the HC$_3$N detections have HCO$^+$/HCN $<1$ and HNC/HCN$>$0.4.}\\ Plots $a$ and $c$ show strong correlations. As HNC/HCO$^+$ increases, galaxies move on the graphs towards higher HNC/HCN and HCO$^+$/HCN ratios. An inverse correlation between HNC and HCO$^+$ line intensities also emerges from plot $b$. Galaxies at HCO$^+$/HCN $<1$ present on average a 30 \% increase in HNC/HCN, compared with those at higher HCO$^+$/HCN ratios. If we exclude the Seyfert NGC~2273, the higher end of the HNC luminosity distribution (see graphs $b$ and $c$ in Fig. \ref{fig:plots}) is composed mainly of luminous infrared galaxies.\\
The general picture emerging from the graphs is that starbursts are characterized by faint HNC and bright HCO$^+$ emission (compared with HCN), while LIRGs mostly occupy the opposite end of the HNC-HCO$^+$ correlation, with high HNC and low HCO$^+$ intensities. All HNC-bright galaxies are either LIRGs or Seyferts. Most of the HC$_3$N detections are LIRGs, the only exception being the Seyfert NGC~1068. 

\subsection{Line ratios at 112 GHz}
\label{sec:112}

The main detections in the 112 GHz band are the $J$=1-0 emission lines of CO, $^{13}$CO and C$^{18}$O, and the spin doublet of CN 1--0 $J$=3/2-1/2 and $J$=1/2-1/2. These lines are detected in most of the galaxies in our sample, and their integrated intensities are compared in the  plots in Fig. \ref{fig:ratplot112}. \\
A particularly rich chemistry is found in IC~860, whose spectrum shows bright emission lines of methanol and HC$_3$N $J$=12--11. This detection of HC$_3$N is the only one in the 112 GHz sample and will be further discussed in section \ref{sec:hc3n_ic860}.

\subsection{Principal component analysis}
\label{sec:pca}

\begin{table}
\begin{center}
\renewcommand{\arraystretch}{1.2}
\setlength{\tabcolsep}{2pt}
{\tiny
\begin{tabular}{l c c c c c c}
\hline
\hline
\\
 & HNC/HCN & HCO$^+$/HCN & C$_2$H/HCN & Silicate & PAH & 25/100 $\mu$m \\
\\
HNC/HCN &    1.00  &  -  &  - & - &  - & -\\
HCO$^+$/HCN &   -0.43  & 1.00   & - &  - &   - &  -\\
C$_2$H/HCN &    0.46  &  0.09  &  1.00  & - &   - &  -\\
Silicate &   -0.11  & -0.37 &   0.03 &   1.00 &  - &   -\\
PAH &   -0.41  &  0.66 &   0.19 &  -0.22 &   1.00 & -\\
25/100 $\mu$m &   -0.24  & -0.10  & -0.2 &   0.03 &  -0.46  &  1.00\\
\\
\hline
\end{tabular}}
\caption{\label{tab:corr}Correlation matrix}

\end{center}
\end{table}

Line-ratio diagrams as the ones shown in sections \ref{sec:88} and \ref{sec:112} represent the standard framework for interpreting molecular data and can be easily compared with previous studies \citep[e.g., ][]{baan08,loenen08let,baan10}, as discussed in section \ref{sec:compa}. In this approach, the properties of different galaxy types are compared two at a time to find the observables that best characterize different physical environments. In general, however, the observed quantities are not independent and different environments are best described by a combination of observables. As the number of observables increases, it becomes more and more difficult to interpret multidimensional datasets by means of 2D sections (i.e., line-ratio diagrams).\\
To address this complexity, we applied a principal component analysis (PCA) to our data. This is commonly used to reduce the dimensionality of a dataset and is very effective in finding hidden trends that could otherwise be buried in the noise. For an application of a PCA to molecular maps, see, e.g.,  \citet{ung97} and \citet{meier05}.\\
Each galaxy in our sample is described by a set of $n$ observed quantities (e.g. line ratios, IR properties) that represent an initial base of vectors. The PCA algorithm first computes the covariance matrix of the data along the $n$ directions and finds its eigenvalues and eigenvectors. The eigenvalues are then sorted in descending order, and the corresponding eigenvectors labelled as {\it principal component (PC) 1, 2, etc...} \\
The first principal component (PC 1) is thus the linear combination of the initial galaxy properties along which the dispersion of the data is maximum. PC 2 is the vector, perpendicular to PC 1, which has the second highest dispersion, and so on for all $n$ PCs. The values of the observed properties for each galaxy are then projected on the new base of PCs.\\
In our analysis we choose as a base of observables the line intensity ratios HNC/HCN, HCO$^+$/HCN, and C$_2$H/HCN and the IR properties silicate absorption, PAH EW and the ratio of the IRAS fluxes at 25 and 100 $\mu m$. Because our algorithm cannot deal with upper limits, we do not include faint lines, as HC$_3$N. In order to maximize the number of galaxies, we limit our analysis to the 88 GHz band, since only a fraction of our sample was observed at 112 GHz.

\begin{table}
\renewcommand{\arraystretch}{1.2}
\begin{center}
{\tiny 
\begin{tabular}{l c c c c c c } 
\hline 
\hline 
\\
 & PC 1 & PC 2 & PC 3 & PC 4 & PC 5 & PC 6 \\
\\
\hline 
\\
Variance \%  &  45  &  29  &  16  &  6  &  3  &  1\\ 
\\
\hline 
\\
HNC/HCN & 0.03  & -0.19  &  0.59  & -0.32  &  0.02 & -0.71\\ 
HCO$^+$/HCN & -0.24  &  0.17  & -0.26  & -0.18  & -0.9  & -0.21\\ 
C$_2$H/HCN   & -0.02  &  0.10  &  0.26  & -0.79  & -0.01  &  0.54\\ 
Silicate  & 0.83  &  0.52  & -0.06  & -0.06  & -0.07  & -0.12\\ 
PAH   & -0.46  &  0.68  & -0.20  & -0.17  &  0.42  & -0.28\\ 
25/100$\mu m$   & 0.17  & -0.43  & -0.69  & -0.45  &  0.22  & -0.24\\ 
\\
\hline
\end{tabular}}
\end{center}

\caption{\label{tab:pca} Projection of the principal components onto the base of observables. For each PC, its contribution to the total variance in the data is also shown.}
\end{table}

\subsubsection{Results of the PC analysis}
\label{sec:pcares}

\begin{figure*}
\begin{centering}
\includegraphics[width=1\textwidth,keepaspectratio]{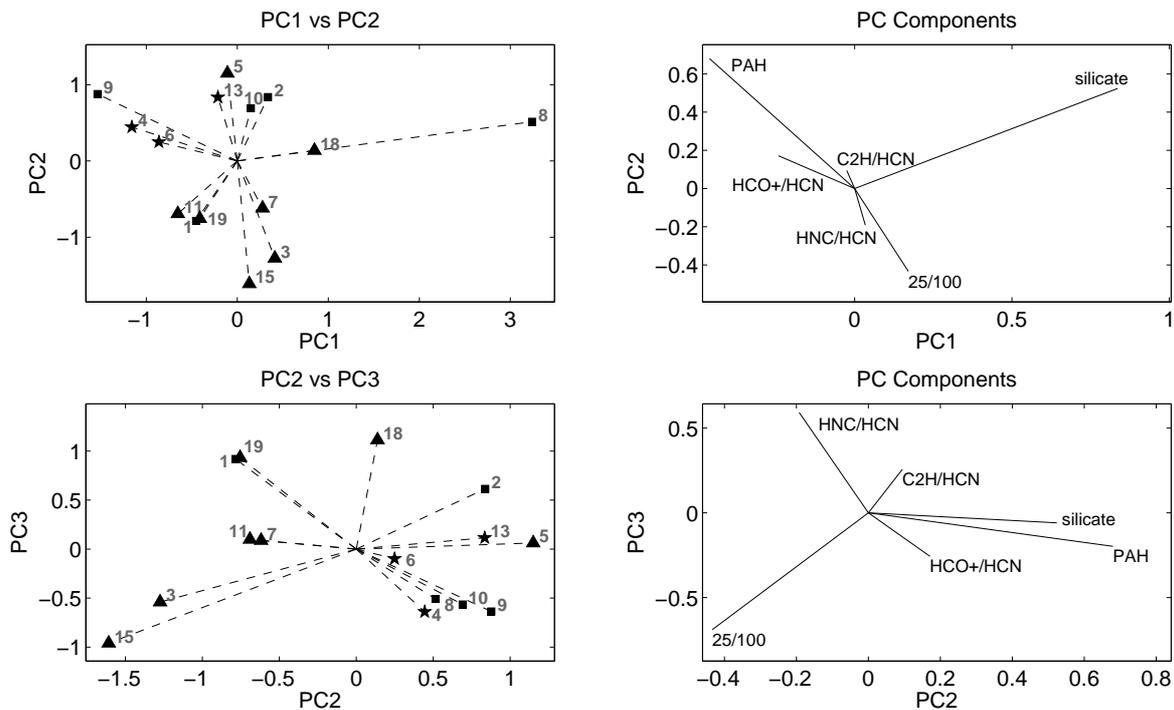}
\caption{\label{fig:pca} Results from the principal component analysis. $Left$: Position of the observed galaxies in the PC base $[PC 1 , PC 2, PC 3]$. Different symbols refer to starbursts ({\it stars}), Seyferts ({\it triangles}) and LIRGs ({\it squares}). Galaxies with both AGN and LIRG properties are plotted as {\it triangles}. Numbers indicate different galaxies, as explained in Table \ref{tab:linrat}. Values for the PAH equivalent widths are taken from \citet{spoon07}. Infrared fluxes were obtained from the NASA/IPAC Extragalactic Database. $Right$: Projection of the PC components onto the original base of observables. We consider only the first three PCs, since they account for 90 \% of the variance and thus contain most of the information in the dataset.}
\end{centering}
\end{figure*}

The results of the PCA algorithm are shown in Table \ref{tab:pca}. Here the projection of the PCs onto the original base are reported along with their contribution to the total dispersion in the dataset. Most of the information is contained in the first three PCs, which account for about 90 \% of the total variance. In the following analysis we will thus focus on PC {\it 1,2,3}.\\
In Fig. \ref{fig:pca} \, $right$ we show the projection of the PCs onto the original base of observables, while in Fig.~\ref{fig:pca} \, $left$ we plot the position of the galaxies in the PC base. \\
The main contribution to the dispersion of the data comes from PAH EW and silicate absorption, which have the highest projections on PC 1. The opposite signs mean that the dispersion along PC1 is mainly caused by galaxies that have low PAH EW and high silicate absorption, or vice versa. The first PC is therefore dominated by the diagonal sequence on the mid-IR diagram of Fig.~\ref{fig:spoondiag}. Galaxies with high values of PC 1 will have high silicate absorption and low PAH EW.
This is evident when comparing Figs.~\ref{fig:spoondiag} and \ref{fig:pca} \, $left$, where the galaxies NGC~4418 and NGC~6090 (8 and 9 on the graphs) lie at the opposite ends of both PC 1 and the mid-IR sequence. \\
Among the molecular line intensity ratios the largest contribution to PC 1 is given by HCO$^+$/HCN. From Fig.~\ref{fig:pca}\, $left$,  HCO$^+$/HCN appears to be positively correlated with PAH EW (same sign of the projection along PC 1). This agrees with Figs.~\ref{fig:plots}$d$ and \ref{fig:hncdep} and generally with the trend that will be discussed in Sect. \ref{sec:hcopah}. \\
The second principal component (PC 2) is also dominated by IR observables, but with different contributions than the ones derived for PC 1. Silicates and PAH projections are now positively correlated, and the ratio IRAS 25/100~$\mu$m gives a significant contribution to the dispersion. As can be seen in Fig.~\ref{fig:pca}\, $left$, {\it PC 2 efficiently separates AGNs from obscured or starburst-dominated sources}. All AGNs, with the exception of NGC~3079 (number 5 on the graph), have negative PC 2 values, which correspond roughly to classes 1A-1B on the mid-IR diagram of Fig. ~\ref{fig:spoondiag}. 
An interesting anti-correlation between PAH EW and IRAS 25/100 $\mu$m flux ratio is also evident in Fig.~\ref{fig:pca}\, $left$. This may be caused by higher dust temperatures in systems with a dominant AGN component.\\
Most of the molecular information is contained in PC 3, where a substantial fraction of the total dispersion is along the HNC/HCN--HCO$^+$/HCN direction, while the contribution by PAH and silicates is almost negligible. The IRAS colour (25/100~$\mu$m ratio) still plays an important role, with a projection along PC 3 ($\simeq$-0.7) slightly higher in absolute value than the one for the HNC/HCN line intensity ratio ($\simeq$0.6). The opposite sign of these two projections is caused by a slight anti-correlation of the HNC/HCN ratio with dust temperature, which also results from the correlation matrix in Table \ref{tab:corr}. This trend, however, is not evident in Fig. \ref{fig:plots}$e$.
The distribution of galaxies along PC 3 does not show any obvious correlation with galaxy properties and its interpretation is not straightforward. We will further discuss the results of the PC analysis in Paper II, when we will have additional information from VLA radio observations about, e.g., gas surface density and star-formation rate.

\subsection{Comparison with previous observations}
\label{sec:compa}
\begin{figure*}
\centering
\includegraphics[width=.45\textwidth,height=.35\textwidth]{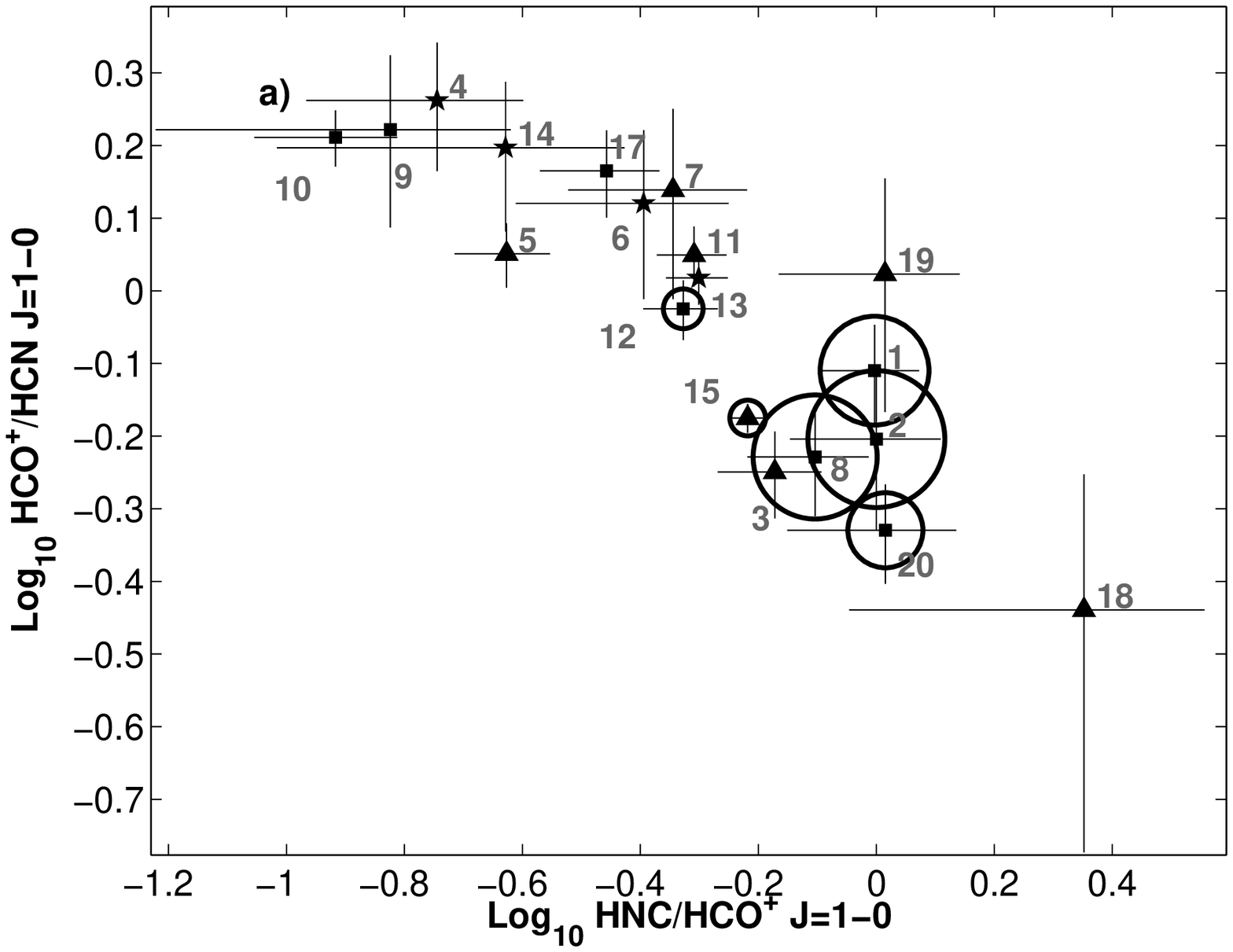}
\includegraphics[width=.45\textwidth,height=.35\textwidth]{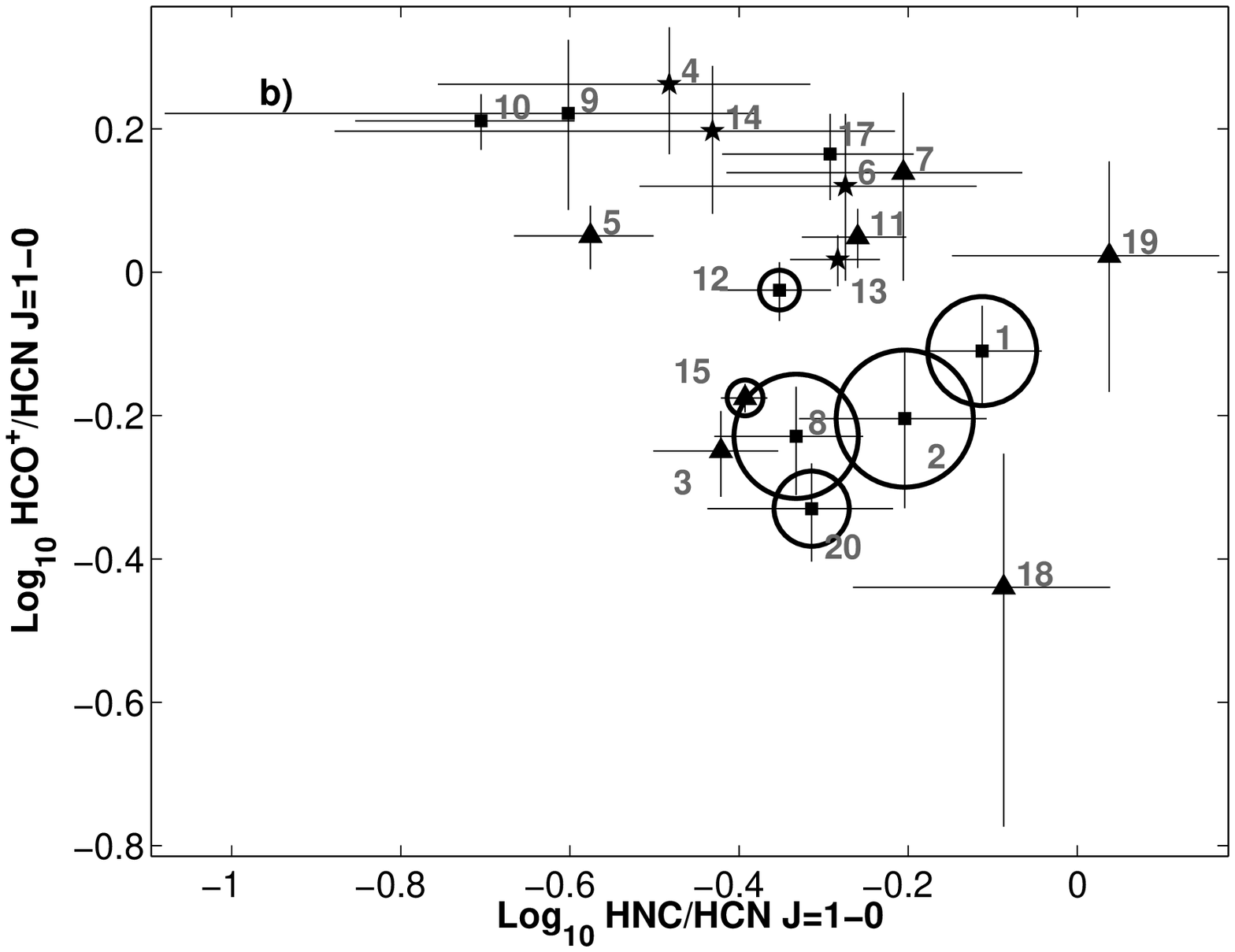}
\includegraphics[width=.45\textwidth,height=.35\textwidth]{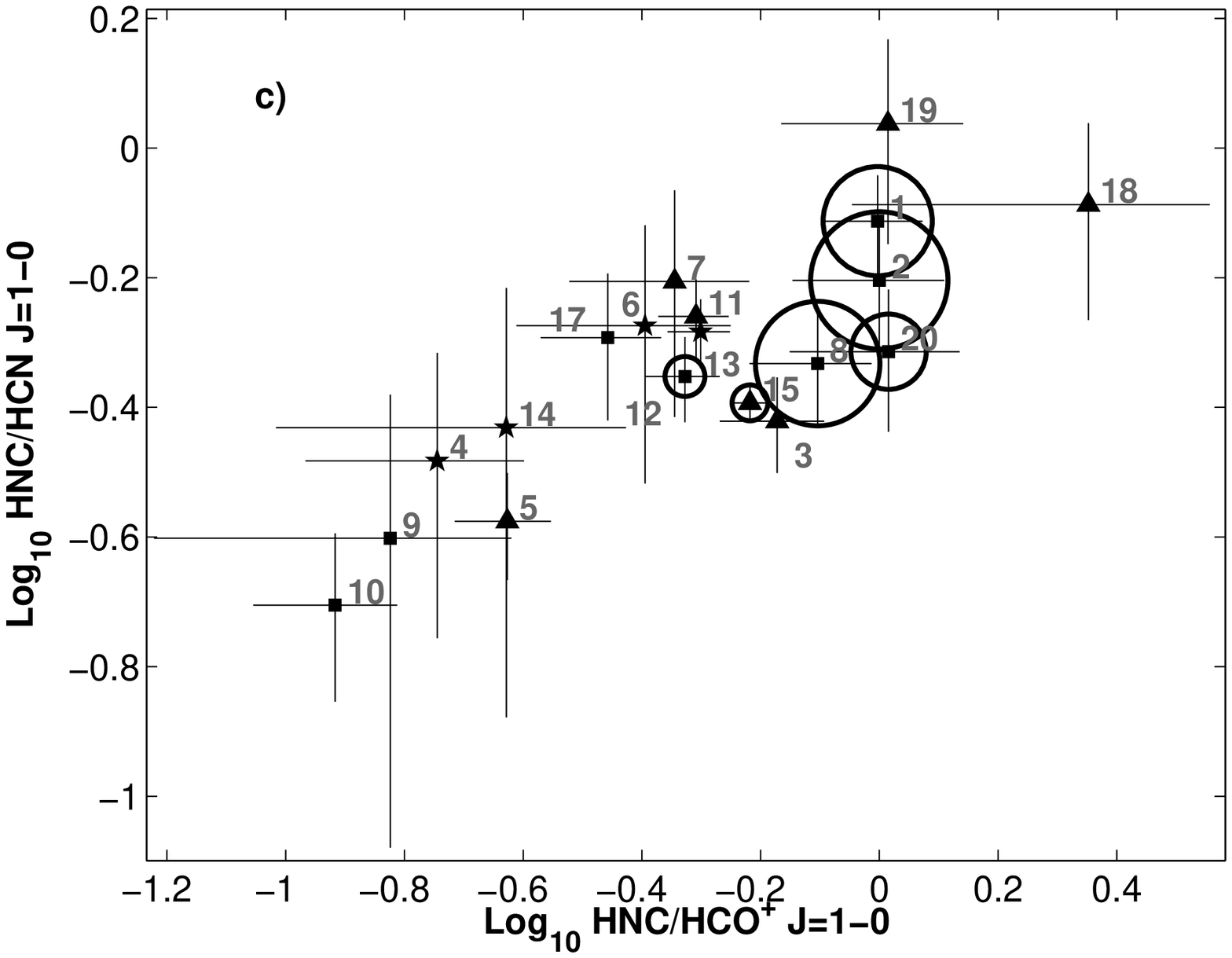}
\includegraphics[width=.45\textwidth,height=.348\textwidth]{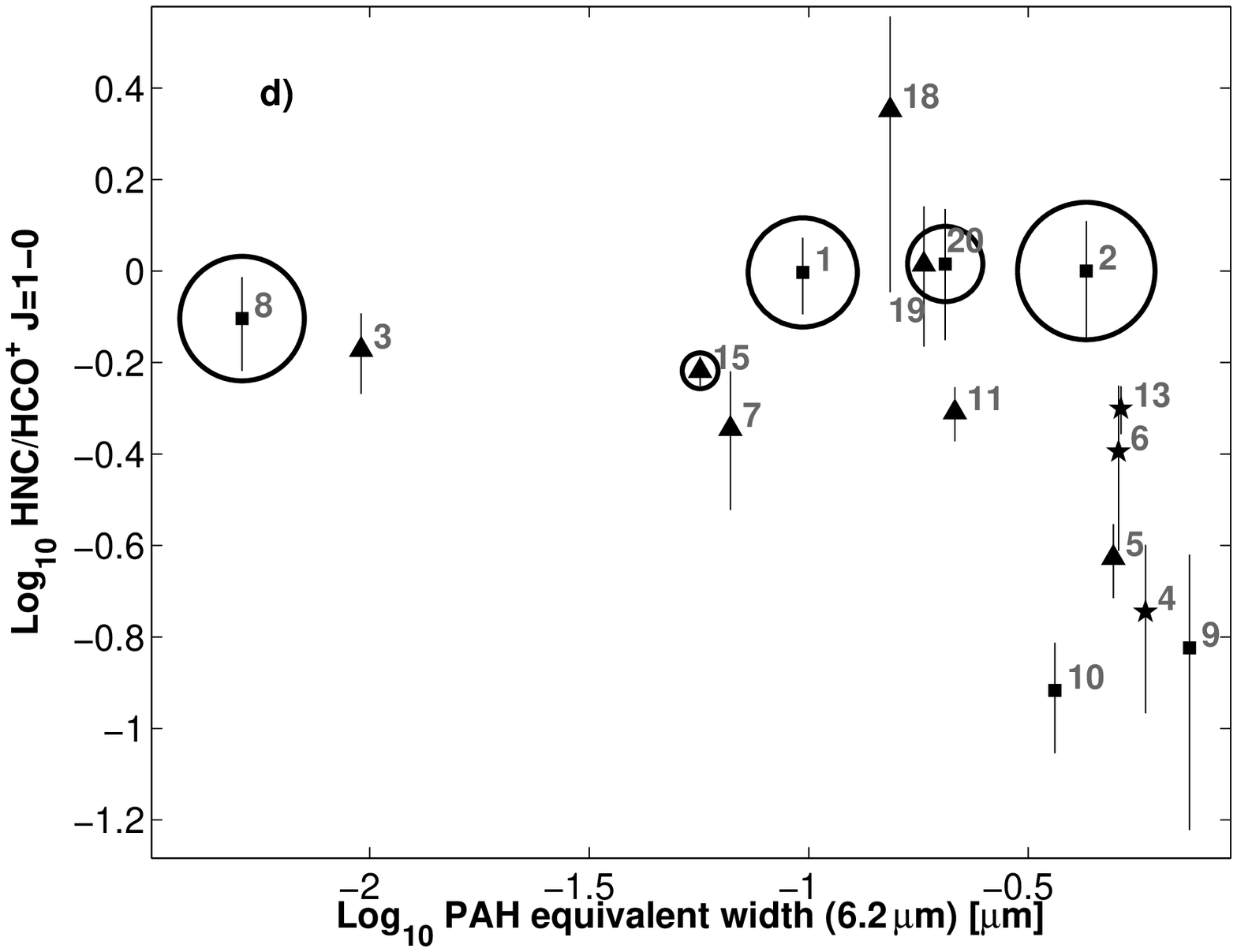}
\includegraphics[width=.45\textwidth,height=.35\textwidth]{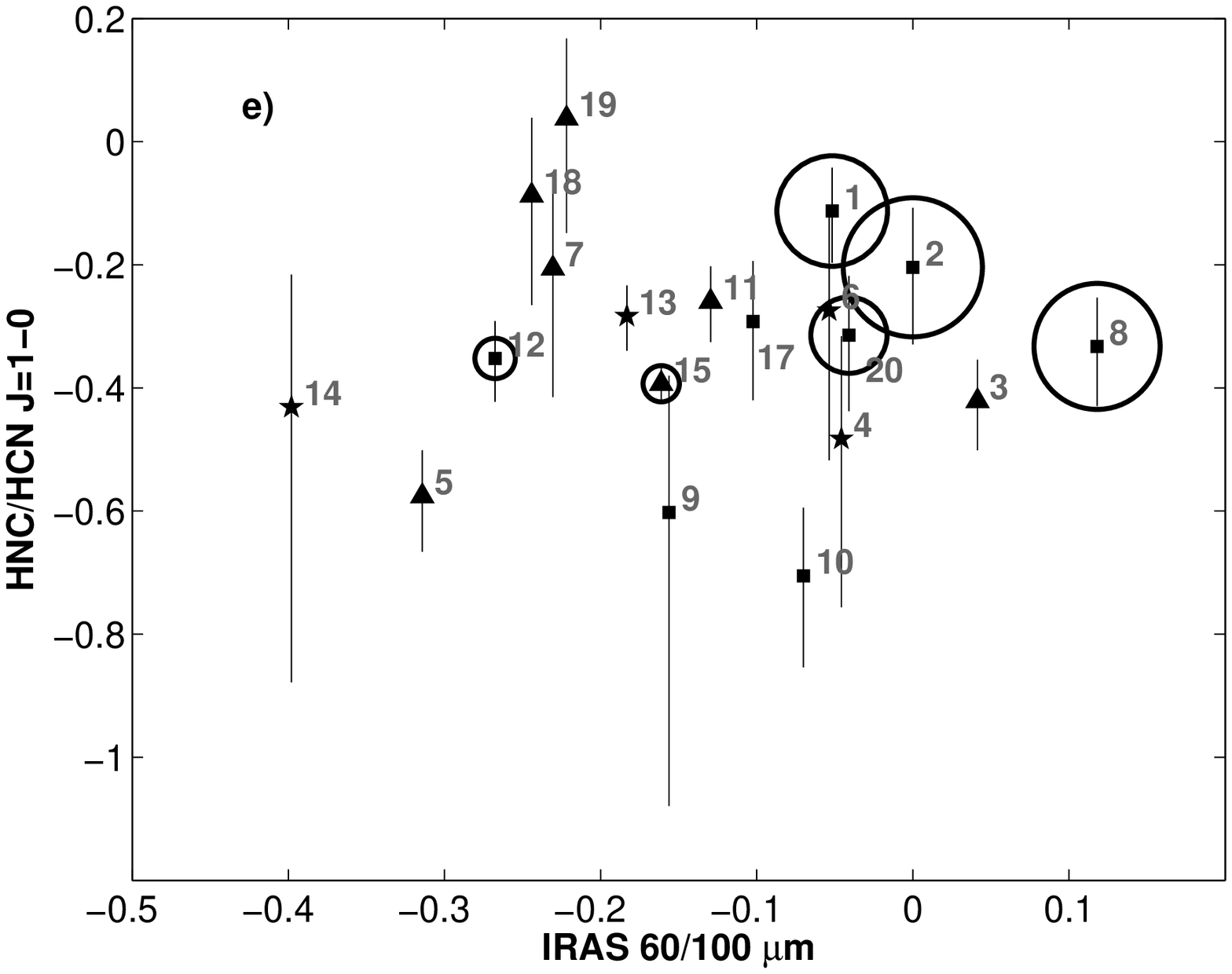}
\includegraphics[width=.45\textwidth,height=.35\textwidth]{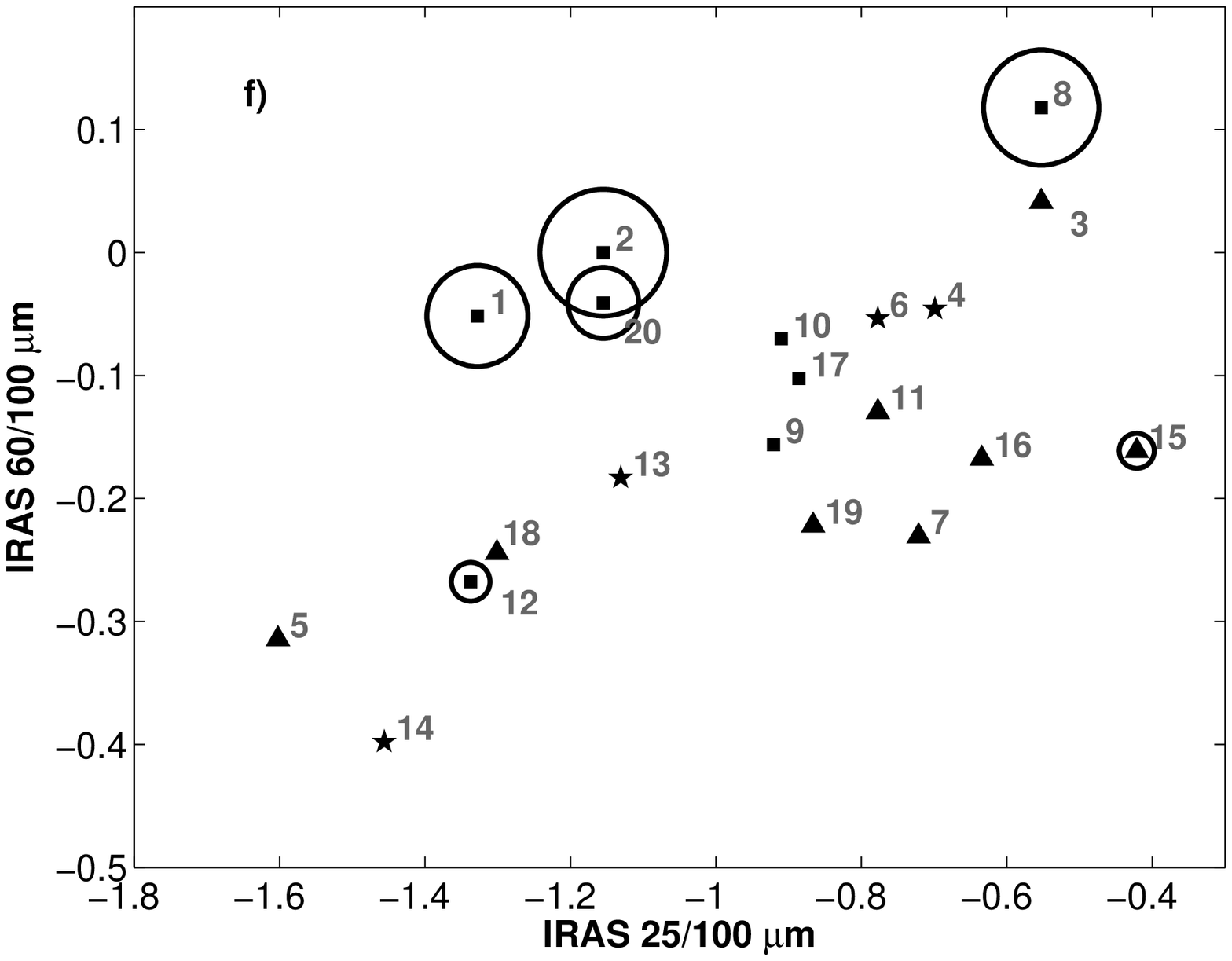}
\caption{\label{fig:plots}Diagrams derived from observed line ratios.  Symbols and galaxy labelling are the same as in Fig. \ref{fig:pca}. Galaxies where HC$_3$N was detected are marked with a circle. The diameter of the circle is proportional to the HC$_3$N 10--9/HCN 1--0 line ratio. Error bars show 1-$\sigma$ uncertainties.}
\end{figure*}

A previous study of the molecular emission of dense gas in luminous infrared galaxies was reported by \citet{baan08}. The authors analyse data for 117 galaxies, with infrared luminosities ranging over about three orders of magnitude. Our diagrams in Fig. \ref{fig:plots} can be directly compared with those of Fig. 9  in \citet{baan08}. \\
The correlations between HNC/HCO$^+$ and the two ratios HNC/HCN and HCO$^+$/HCN, shown by  \citet{baan08} in plots $a$ and $c$, are confirmed by our observations. Our measurement have a much smaller scatter, thanks to the excellent relative calibration of the line intensities provided by the EMIR receiver. However, we cannot exclude that the smaller scatter is partially caused by source selection effects.\\
Plot $b$ in \citet{baan08}, showing  HCO$^+$/HCN versus HNC/HCN, however, cannot be reproduced by our data. In particular, we did not observe any galaxy with $\log$ HNC/HCN$<$-0.4 and $\log$ HCO$^+$/HCN$<$0, while about 25 \% of the sources plotted by \citet{baan08} are in this range. \\
As a result, the correlation between HCO$^+$/HCN and HNC/HCN line ratios in our graph $b$ goes in the opposite direction, with low HNC/HCN corresponding to high HCO$^+$/HCN. It is not clear whether this is caused by selection effects that may affect our observations, or to the large scatter in the data in \citet{baan08}.\\
In Table \ref{tab:comp} we compare line intensities from \citet{baan08} with our EMIR values. In most cases the two datasets agree at the 20 \% level, with some exceptions. The most striking differences are seen for HNC/HCN in NGC~7469 and HCO$^+$/HCN in NGC~6240, which vary more than 60 \%, passing from values higher than unity in one dataset to much lower ratios in the other. \\
These discrepancies could be caused by calibration or pointing inaccuracies, which affect most narrow band single dish observations. The determination of line ratios with EMIR is more robust, because these inaccuracies do not affect the relative intensity of lines detected in the same band. 
\section{Discussion}
\label{sec:discussion}

\subsection{Interpreting line-intensity ratios}

Line-intensity ratios are often interpreted as estimates of the abundance ratios of two molecular species. This is true only if the emission is optically thin and the two molecules have similar excitation properties. Moreover, single dish observations only reveal global ratios, averaged on the whole galaxy, and the interpretation of the results has to take into account the lack of spacial information.\\ 
In our study we detect transitions with critical densities ($n_\mathrm{c}$) that vary by almost four orders of magnitude (see Table \ref{tab:bands}), and it is therefore likely that excitation plays an important role in determining the observed ratios. \\
Transitions with large $n_\mathrm{c}$ are efficiently excited in high density gas, which is mostly concentrated in the inner parts of galaxies, while the emission of lines with lower $n_\mathrm{c}$, which can be excited in more diffuse gas, is generally more extended. Low-$n_\mathrm{c}$ lines will thus have a higher beam filling factor, resulting in an increased T$_\mathrm{A}^\star$. In this context, it is advisable to compare line intensities of transitions with similar $n_\mathrm{c}$, since the resulting ratios can be more directly related to molecular abundances. \\
Also, the excitation temperature of the transitions is an important factor when interpreting single dish intensities. If the emission is coming from regions with different gas temperatures, this translates into a different population of the low-J levels, which directly affects the line intensity ratios.\\
Another complication arises when we take into account radiative excitation. Many of the observed molecules, as e.g. HNC, HCN, and HC$_3$N, can in fact be affected by radiative pumping via IR vibrational modes. Radiative pumping of HNC has been proposed by \citet{aalto_07} as an explanation of the high HNC/HCN ratio observed in LIRGs, while the strong connection of HC$_3$N excitation with the IR continuum has been discussed in \citet{costagliola2010}. \\
Extreme caution should be used therefore when interpreting global line intensity ratios as abundance indicators. Molecular excitation across our galaxy sample will be further discussed in a following paper, where we will include 1 mm counterparts of the observed lines.

\subsection{Is HCO$^+$/HCN driven by XDR chemistry?}
\label{sec:hco+}
The HCO$^+$/HCN $J$=1-0 line ratio has been proposed as a reliable diagnostic tool to distinguish between AGN- and starburst-powered galaxies \citep[e.g., ][]{kohno2001,imanishi04,imanishi07}. Interferometric observations by \citet{kohno2001} reveal that Seyfert galaxies have lower  HCO$^+$/HCN ratios compared with starbursts. Seyfert galaxies with starburst-like HCO$^+$/HCN ratios are generally interpreted as mixed AGN-starburst objects \citep{imanishi07}. 
In our observations, starbursts and LIRGs have, on average, higher HCO$^+$/HCN ratios than AGN, which is consistent with the trend observed by \citet{kohno2001} and \citet{imanishi07}. This is commonly attributed to an enhancement of HCN abundance in the X-ray-dominated region \citep[XDR, ][]{maloney96} surrounding an AGN \citep{lepp96}. Chemical models by \citet{meijerink05} show that an HCO$^+$/HCN line ratio lower than one can be observed on the surface of low-density ($n<$10$^5$ cm$^{-3}$) XDRs, for H$_2$ column densities $<$10$^{22}$ cm$^{-2}$. However, at higher depths inside the molecular cloud,  HCO$^+$ abundance increases and eventually the cumulative HCO$^+$/HCN column density ratio becomes higher than one. \citet{meijerink07} report that for N(H$_2$)$>$10$^{23}$  cm$^{-2}$, the intensity of the $J$=1--0 and $J$=4--3 transitions emerging from an XDR is higher for HCO$^+$ than for HCN by a factor of at least three. This is attributed to the fact that HCN abundance in an XDR is enhanced (compared with HCO$^+$) only in a narrow range of ionization rate/density \citep[Fig. 3 in ][]{lepp96}. \citet{meijerink07} conclude that a low HCO$^+$/HCN line ratio is a good tracer of UV-dominated regions (PDR) for gas densities $>$10$^5$ cm$^{-3}$, rather than X-ray-irradiated gas. {Thus the observed low HCO$^+$/HCN line ratio could be either caused by low-density ($n<$10$^5$ cm$^{-3}$) XDRs or dense ($n>$10$^5$ cm$^{-3}$) PDRs.}\\ Observations by \citet{blake87} in the Orion molecular cloud reveal extremely low HCO$^+$/HCN abundance ratios ($<$10$^{-3}$) in {\it hot cores}, i.e. warm and dense gas around massive young stars. This is supported by calculations by \citet{bayet08}, who find abundances of HCO$^+$ as low as 10$^{-12}$ (relative to H$_2$) in a wide range of hot core models.  The faint HCO$^+$ emission may thus be caused by deeply embedded star-formation, instead of XDR or PDR chemistry. \\
This interpretation is based on the assumption that line ratios are directly linked to relative abundances. This may not be true
for HCN and HCO$^+$, which have different excitation properties. The critical density of the $J$=1--0 transition of HCO$^+$ is one order
of magnitude lower than for HCN. As a consequence, HCO$^+$ emission can originate from lower density gas, which in general has a
higher filling factor compared with the dense ($n_\mathrm{H}>$10$^5$ cm$^{-3}$) component. This implies that the high HCO$^+$/HCN
line ratio observed in starburst galaxies may be caused by different filling factors for the two molecules instead of by their
abundance. \\
Although \citet{krips08} suggest that the HCN/HCO$^+$ line ratio can often serve as a measure of abundance, we advise caution in doing so. Recent work by \citet{sakamoto09} and \citet{aalto09} show that  in luminous, dusty nuclei absorption and radiative excitation affect the line emission of these molecules on global scales.

\subsection{The HNC/HCN ratio}
\label{sec:hnc} 
In the Galaxy, the HNC/HCN observed line ratio ranges from 1/100 in hot cores \citep{schilke92} to values as high as 4 in dark clouds \citep{hirota98}. The abundance of the HNC molecule decreases with increasing gas temperature. \citet{hirota98} suggest that this may be owing to the temperature dependence of neutral-neutral reactions, which, for temperatures exceeding 24 K, selectively destroy HNC in favour of HCN. Interestingly, \citet{greaves96} find HCN/HNC abundance ratios of up to 6 for Galactic spiral arm clouds (GMCs). The highest ratios are observed in regions of lowest N(H$_2$), consistent with models by \citet{schilke92}, which show and increase of HCN/HNC with increasing C/CO. Bright HNC emission is commonly observed in extragalactic objects \citep{aalto02,wang04,meier05,perez07}. In particular, overluminous HNC $J$=3--2 is found in LIRGs \citep[e.g., ][]{aalto_07}, where gas temperatures, derived by the IR dust continuum \citep{evans03} and mid-IR molecular absorption \citep{lahuis07}, are usually $>$ 50 K and can reach values as high as a few 100 K. Here, ion-molecule chemistry in PDRs may be responsible for the observed ratios. \citet{meijerink07} find that the HNC/HCN $J$=1--0 line ratio is enhanced in PDRs and can reach a maximum value of one for H$_2$ column densities exceeding 10$^{22}$ cm$^{-2}$. Even higher HNC/HCN line ratios can result from XDR emission. Models by \citet{meijerink07} show that for the $J$=4--3 transition, this ratio can be as high as 1.6 in dense (n(H)$>$10$^6$ cm$^{-3}$) X-ray-dominated regions. The average HNC/HCN $J$=1--0 line ratio for AGN and LIRGs in our sample is $\simeq$0.5 (see plot $c$ in Fig. \ref{fig:plots}), which is consistent with the results by \citet{meijerink07}  for XDRs with densities $<$10$^5$ cm$^{-3}$ and high X-ray fluxes, F$_X>$10 erg cm$^{-2}$ s$^{-1}$. For the same sources, low-density XDR chemistry could also explain the observed HCO$^+$/HCN ratios. The HNC/HCN ratio of order unity observed in UGC~5101 and NGC~2273 (see Table \ref{tab:linrat}) is consistent with the emission coming  from either a PDR or a low-density XDR ($n_\mathrm{H}<$10$^5$ cm$^{-3}$). Models of PDRs with different density and radiation field strength can explain HNC/HCN $J$=1-0 line ratios ranging from 0.5 to 1 \citep{meijerink05,loenen08let}. In our sample, the observed HNC/HCN ratio reaches values as low as 0.2 (plots $b$, $c$ in Fig. \ref{fig:plots}), which cannot be explained by existing PDR or XDR models. However, line ratios HNC/HCN$<$0.5 are observed in Galactic PDRs. Abundance ratios of about 0.2 are found, e.g. ,  by \cite{fuente93} in the PDR region of the reflection nebula NGC7023. Low values of HNC/HCN are also found in other galaxies (see , e.g., \citet{baan08}) as discussed in Section  \ref{sec:compa}.

\begin{figure*}
\begin{centering}
\includegraphics[width=.45\textwidth,height=.352\textwidth]{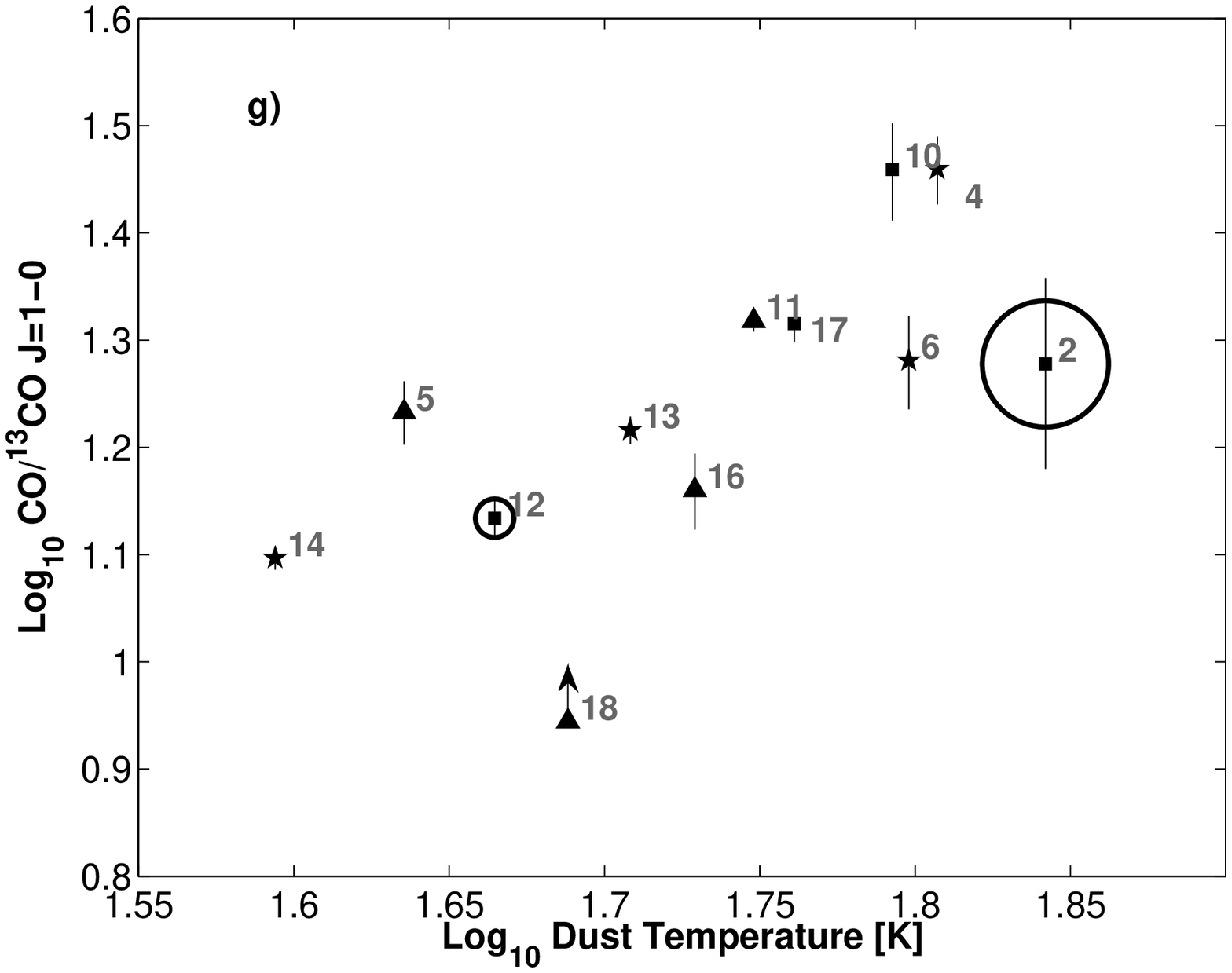}
\includegraphics[width=.45\textwidth,height=.35\textwidth]{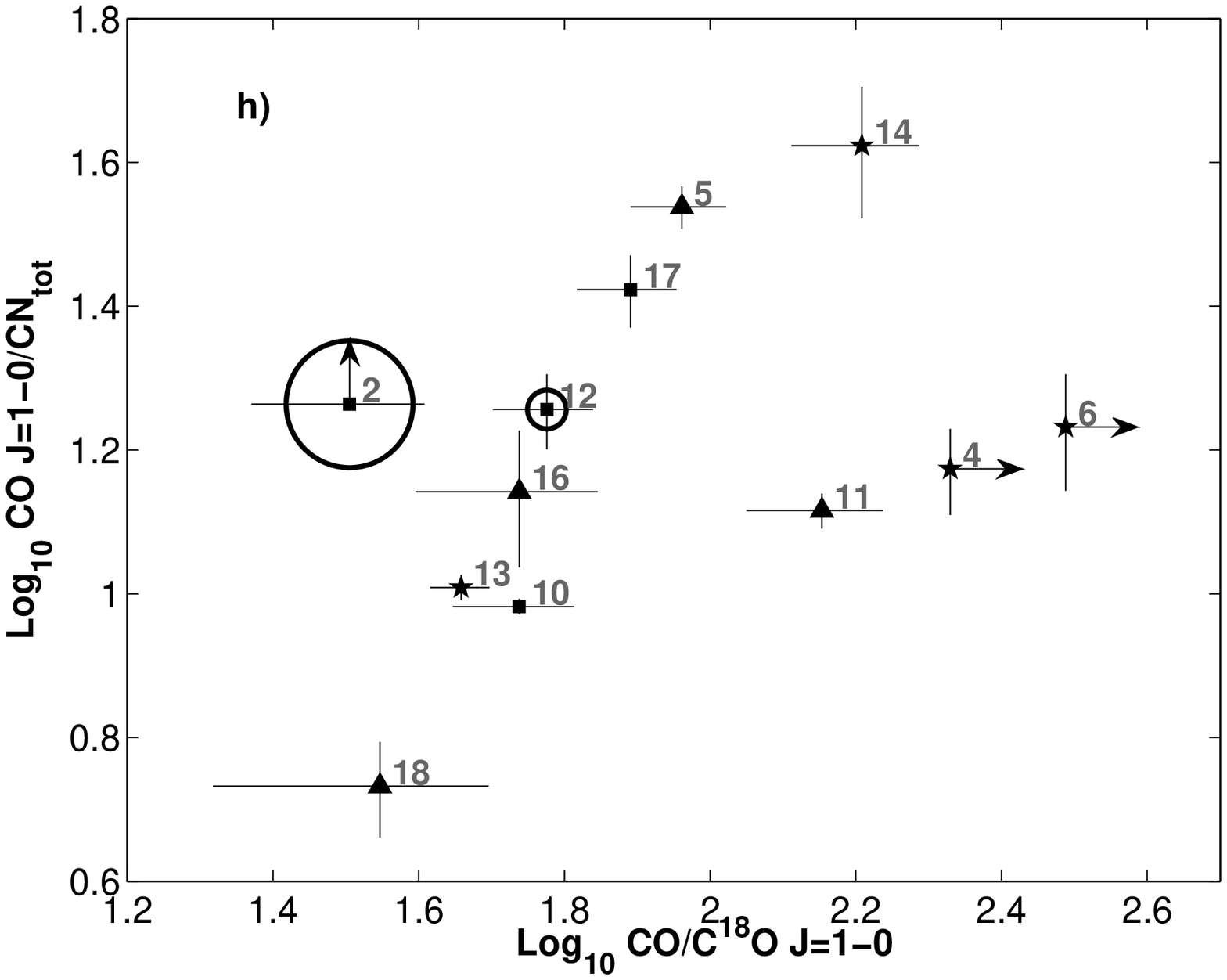}
\caption{\label{fig:ratplot112} See caption of Fig. \ref{fig:plots} and discussion in Sections \ref{sec:cotemp} and \ref{sec:cocn}.}
\end{centering}
\end{figure*}

\subsection{Luminous HC$_3$N: a nascent starburst tracer?}

Bright HC$_3$N emission is often observed in Galactic hot cores \citep{rodriguez98,devicente00}. The molecule is easily destroyed by UV radiation and reactions with C$^+$ and He$^+$ \citep[e.g., ][]{turner98} and can only survive in regions shielded by large gas and dust columns. In a high-resolution study of the chemistry of IC~342, \citet{meier05} find that HC$_3$N emission follows the 3 mm continuum dust emission and anti-correlates with regions of intense UV radiation, such as PDRs. Observations in the Galaxy \citep{wyr99} and in the LIRG NGC~4418 \citep{aalto_07,costagliola2010} show that HC$_3$N is strongly connected with the IR field by means of its vibrational bending modes. In our sample, only five galaxies have detectable HC$_3$N emission: NGC~4418, IRAS~17208, IC~860, NGC~1068,  and NGC~7771. For completeness, we also include in our analysis the detection of HC$_3$N $J$=10--9 in the ULIRG Arp~220 by \citet{aalto02}. Since these objects span a wide range of LSR velocities, in most cases the detectability of HC$_3$N does not depend on the  sensitivity of our measurements, but on intrinsic properties of the observed galaxies. This is further discussed in Appendix \ref{sec:dethc3n}, where we compare HCN and HC$_3$N detection thresholds. \\
All galaxies with detected HC$_3$N have HCO$^+$/HCN $<$1 and higher-than-average  HNC/HCN ratios. This is consistent with the hot core models described in \citet{bayet08}, as discussed in Sect. \ref{sec:hco+}. It is thus possible that the bulk of the molecular emission in these objects comes from regions of young star-formation. This interpretation was already proposed by \citet{aalto_07} to explain the HC$_3$N emission of NGC~4418 in terms of a {\it nascent starburst}. \\
The HC$_3$N-bright galaxies reside in the HCO$^+$/HCN interval that in the diagnostic diagrams discussed by \citet{kohno2001} is associated with Seyfert galaxies. \citet{kohno2001} and \citet{imanishi07} interpret the observed ratios as the signature of XDR chemistry driven by the hard radiation from the AGN. Our data seem to challenge this interpretation, since HC$_3$N cannot survive in significant abundance in highly irradiated gas. Bright HC$_3$N emission can emerge from an AGN only if the central source is shielded by a considerable amount of gas and dust. In these objects, the bulk of the molecular emission would be coming from an hot core-like region instead of being dominated by XDR chemistry. Plots $e$ and $f$ in Fig. \ref{fig:plots} also show that the galaxies with the highest HC$_3$N/HCN ratios have high ratios of IRAS 60 to 100 $\mu$m fluxes. This may indicate that HC$_3$N is more abundant in warm environments.\\ An enhanced HC$_3$N emission at high gas temperatures could also be owing to excitation effects. The upper state energy of the observed HC$_3$N transitions is indeed of the order of 20-30 K, higher by a factor of  $\simeq$5 than typical energies for other high density tracers, meaning that the molecule is more efficiently excited at higher temperatures. The bright HC$_3$N emission is thus likely to be a combination of enhanced abundance and favourable excitation conditions in warm, dense gas.\\
Radiative pumping of the molecule's rotational levels by means of IR vibrational modes was also suggested by \citet{costagliola2010}.

\subsection{CO/$^{13}$CO vs dust temperature}
\label{sec:cotemp}

Plot $g$ in Fig. \ref{fig:ratplot112} shows how the ratio $\Re_{1-0}\equiv$CO/$^{13}$CO $J$=1--0 varies with dust temperature. The dust temperature was calculated from IRAS fluxes at 100 and 60 $\mu$m, via the formula
\begin{equation}
T_\mathrm{dust}=-(1+z)\left[ \frac{82}{ln\left( 0.3 f_\mathrm{60\mu m}/f_\mathrm{100\mu m}\right)}-0.5\right],
\end{equation}
see, e.g., \citet{solomon97}. The IRAS fluxes were obtained from the NASA/IPAC Extragalactic Database.\\
A clear trend is observed, with $\Re_{1-0}$ increasing as $T_\mathrm{dust}$ increases. This is a well established result, see, e.g., \citet{young86} and \citet{aalto95} for a detailed discussion.
The trend is generally attributed to opacity effects arising in a collection of non-uniform clouds, averaged in the beam. In a molecular cloud, the emission of CO 1-0 arises mostly from the outer layers, up to optical depths $\tau\simeq$1, while the $^{13}$CO emission stays in general optically thin deeper into the volume of the cloud. The line ratio $\Re_{1-0}$ is thus mostly affected by the relative opacity of the two species.\\
The opacity of the CO 1-0 line strongly depends on the excitation temperature \citep[see, e.g., ][ Eq. 1]{aalto95}. With rising T$_\mathrm{ex}$, the $J$=1 level is in fact de-populated, causing a decrease in optical depth of the CO 1-0 transition. Where $\tau_\mathrm{CO}\simeq$1, we can assume $\tau_\mathrm{^{13}CO}<<$1 and the intensity of the $^{13}$CO $J$=1--0 line is an even steeper function of T$_\mathrm{ex}$ than the $^{12}$CO intensity \citep[see, e.g., ][]{mauer93}. As a consequence, $\Re_{1-0}$ must increase with temperature.\\
In our Galaxy \citet{young86} find that $\Re_{1-0}$ has an average value of roughly 8, while, in extragalactic surveys, values $\Re_{1-0}>$20 are observed \citep[e.g., ][ this work]{aalto95}, the highest ratios being found in interacting systems and mergers. 
In our sample, the highest $\Re_{1-0}\simeq28$ is found for the galaxies NGC~6240 and NGC~1614. In these objects, high gas temperatures and a diffuse ISM may result in an extremely low optical depth of the CO 1--0 line, which would explain the observed ratios.\\
Additional processes have been proposed in order to explain these extreme ratios. 
In the central regions of merging and interacting galaxies, large amounts of gas and dust are trapped in the deep potential well, giving rise to high gas pressures. Large turbulent motions, powered by differential rotation or stellar winds from an enhanced star-formation are required to sustain the clouds against collapse. This would cause a broadening of the CO emission line and a decrease in optical depth, with a consequent increase in $\Re_{1-0}$ \citep{aalto95}.\\
Chemistry could also provide an explanation of the observed high ratios. In star-forming regions the surfaces of molecular clouds are indeed dominated by UV-photons from young stars. In these regions, $^{13}$CO is dissociated more easily than CO, which is less affected by the UV radiation thanks to self-shielding \citep{aalto95}.  This process could be relevant in starburst galaxies, as  NGC~6240 and NGC~1614 in our sample.\\
The age of the starburst could also have an influence on $\Re_{1-0}$. Nuclear processing in stars favours $^{13}$CO over CO, and the evidence for an enhanced  $\Re_{1-0}$ at low metallicities has been reported by, e.g., \citet{young86}. A large $\Re_{1-0}$ would be thus compatible with
a young star-formation age.

\subsection{CO/CN vs CO/C$^{18}$O}
\label{sec:cocn}

Plot $h$ in Fig. \ref{fig:ratplot112} shows the ratio  CO/CN vs CO/C$^{18}$O. The total CN intensity was obtained by adding the integrated intensities of the two lines of the spin doublet $J$=3/2-1/2 and $J$=1/2-1/2. The ratio CO/CN varies in our sample over roughly one order of magnitude and shows a general trend of increasing CO/CN with increasing CO/C$^{18}$O. This implies that CN and C$^{18}$O emission are positively correlated.
There are several potential explanations of this:
\begin{itemize}
\item We can assume that both C$^{18}$O and CN are optically thin, for CN we know this from the CN 1--0 spin group ratio, which is at the optically thin level for all  detected galaxies (apart from, possibly, UGC~5101). In this case, the intensity of CN and C$^{18}$O would scale with the total molecular column density. This linear scaling with total column would not happen for the CO 1--0 intensity, which can be considered to be at least moderately optically thick in all the galaxies in our sample. The observed correlation would then be a result of optical depth effects.  
\item It is also possible to speculate on other explanations. It is not immediately self-evident that the C$^{18}$O and CN column densities should follow each other. In PDRs, the CN molecule is usually abundant up to a visual extinction A$_V$=2. At higher depths in the clouds the radical CN may react with other molecules to form -- for example -- HCN. Thus, CN can, to some degree, trace the amount of PDR surfaces. \\ In PDRs, it is likely that C$^{18}$O could become photo-dissociated. However, if the CN is tracing PDRs, it is also tracing the impact of the star-formation on host clouds. These are the regions where the ISM enrichment caused by the starburst hits first., i.e.,  where C$^{18}$O would become enhanced early on. Hence, in this scenario, a C$^{18}$O--CN correlation would be expected as a result of ISM enrichment from the starburst. 
\end{itemize} 
In this context, it is interesting to note the distinct deviations from this correlation. We have the three outliers with faint C$^{18}$O compared with their CN luminosities: NGC~4194, NGC~1614 and NGC~7469. A possible way to obtain this comparable overluminosity in CN is to have CN  emerging from an AGN component, with no associated enrichment from a starburst component. \\
The C$^{18}$O enrichment should come quickly in a starburst. If $^{18}$O is a product of $\alpha$-capture on $^{14}$N during He burning, it will come from mass loss in He-burning systems, such as Wolf-Rayet stars and SNe type II. This means that it should come up with prompt nucleosynthesis -- e.g.,  follow O abundance. \\
Because the molecular emission from NGC~7469 is dominated by the starburst ring, this comparatively C$^{18}$O weakness is somewhat puzzling and should be further investigated. The observed high CO/C$^{18}$O ratio could also be caused by a low optical depth of the CO 1--0 line. \\
The other two outliers are different since they are both mergers where the CO emission is dominated by a prominent crossing dust lane – and not the starburst region \citep{olsson2010,aalto2000}. It is therefore possible that the molecular gas in the IRAM beam is dominated by in-falling gas in the dust lane, which has not yet been enriched by the starburst. Indeed, \citet{martin10} find high $^{13}$C/C ratios from CO and CCH observations in the starbursts M~82 and NGC~253, suggesting that the bulk of the molecular mass in these galaxies mostly consists of unprocessed material.\\
An outrigger in a different direction is IC~860. This galaxy has the lowest CO/C$^{18}$O line ratio in the sample -- but the highest CO/CN 1-0 line ratio. In this source, it is possible that only a small fraction of the cloud surface area is in the form of PDRs -- and/or that the CN is being used up in forming HC$_3$N together with C$_2$H$_2$, which is boiling off the grains. Its luminous PAH emission (see Fig. \ref{fig:plots}$d$) does suggest however that PDRs should be quite abundant. It is possible that the PDRs in this object do not coexist with regions dense enough to excite CN emission.

\subsection{Other isotopic line ratios}

For a small subset of galaxies, it was possible to obtain isotopic line ratios for HCN and NCO$^+$, which are reported in Table \ref{tab:iso}. If we assume a Galactic C/$^{13}$C abundance ratio of $\simeq$50 \citep[see, e.g., ][]{wilson94},  or as high as ~100 \citep{martin10}, it is evident that in the galaxies where  $^{13}$C variants were detected, HCN and HCO$^+$ $J$=1--0 transitions are at least moderately optically thick, with  HC$\star$/H$^{13}$C$\star$ as low as 4. 
These data show how the assumption of optically thin lines when interpreting extragalactic molecular intensity ratios is not always correct and opacity effects have to be included in the analysis, even for species much less abundant than CO. The broad band of EMIR allows us to determine isotopic intensity ratios with unprecedented accuracy, which will be of great importance in constraining models of galactic emission. However, deeper integrations are still needed to set reliable limits in more optically thin environments. We also advise caution when interpreting HCO$^+$/H$^{13}$CO$^+$ intensity ratios, since the H$^{13}$CO$^+$ $J$=1--0 line is blended with SiO emission and the two lines are not easy to separate at low SNRs.

\begin{table}[h]
\renewcommand{\arraystretch}{1.2}
\begin{center}
\begin{tabular}{l c c}
\hline
\hline
Galaxy & HCN/H$^{13}$CN & HCO$^+$/H$^{13}$CO$^+$\\
NGC~1614 & $>$2.8 & 4(2) \\
Mrk~231 & 6(2) & 7(6) \\
NGC~4418 & 8(3) & $>$4 \\
NGC~1068 & 17(7) & 24(20) \\
\hline
\end{tabular}
\end{center}

\caption{\label{tab:iso} Isotopic line intensity ratios for HCN and NCO$^+$. Uncertainties (1-$\sigma$) are shown in parenthesis.}
\end{table}

\subsection{HC$_3$N in IC~860}
\label{sec:hc3n_ic860}

In total we detected four emission lines of HC$_3$N in IC~860, reported in Table \ref{tab:ic860}. In the 112 GHz band, we detected two transitions of the vibrationally excited state v$_7$=1 ($e$,$f$), together with the $J$=12--11, v=0 line. In the 88 GHz band we detected the $J$=10--9, v=0 line, and we have a tentative detection of the v$_7$=1$f$, vibrationally excited transition.\\
The interaction between the bending angular momentum of the vibrationally excited states and the rotational angular momentum of the molecule leads to a l-splitting of the levels. Each vibrational state is thus split into two levels, labelled $e$ or $f$ depending on the wavefunction’s parity properties. These parity labels are also shown in Table \ref{tab:ic860}. A useful reference for the labelling of doubled levels in linear molecules is \citet{brown75}.\\
Both $e$ and $f$ states have similar excitation properties and should result in an emission doublet with comparable line intensities. In our spectra, this is true for the $J$=12--11 v$_7$=1$e$ and v$_7$=1$f$ lines at 109 GHz, which show integrated line intensities within the measurement errors. The $J$=10--9 v$_7$=1$f$ transition at 91 GHz has no detected $e$ counterpart and will therefore be treated as an upper limit in our excitation analysis. \\
A first analysis of the excitation of the molecule was performed via the population diagram method. This results in an excitation temperature of 15 K for the v=0 transitions and 42 K for the v$_7$=1 lines.\\
A rotational temperature of 15 K is significantly lower than the 30 K derived by \cite{costagliola2010} for the other HC$_3$N-bright LIRG, NGC~4418. This may suggest that bright HC$_3$N emission could originate from different environments. However, our estimate is based only on two transitions and has to be further investigated. \\
The vibrational excitation of HC$_3$N is strongly dependent on the IR field, since the molecule's bending modes can only be excited by radiation. A higher excitation of the v$_7$=1 lines, compared with the v=0 levels, is therefore expected in a source with a strong IR emission such as IC~860. The derived 42 K is a lower limit to the excitation temperature, because we only have an upper limit for the $J$=10--9 v$_7$=1$f$ transition at 91 GHz.\\
In highly obscured galaxies, the detected IR emission is in most of the cases optically thick. Because millimeter radiation can penetrate a larger column of gas and dust, vibrationally excited HC$_3$N is a valuable probe of the IR field beyond its optically thick surface.\\
The first multi-transition analysis of extragalactic vibrationally excited HC$_3$N was reported by \cite{costagliola2010} for the IR galaxy NGC~4418. Here the authors find a HC$_3$N vibrational temperature of 500 K, much higher than the temperature of the optically thick dust (80 K, \citet{evans03}), which may be the evidence of an obscured compact object in the centre of the galaxy. \\
It would be interesting to perform this kind of analysis on IC~860, which presents unresolved emission in both IR and radio and would then be a good candidate for a deeply embedded compact source.

\begin{table}[h]
\renewcommand{\arraystretch}{1.2}
\setlength{\tabcolsep}{6pt}
\begin{center}
{\small
\begin{tabular}{l c c c c}
\hline
\\
Transition & $\nu$ &  $\int T^{\star}_A dv$ & $\Delta v$  & E$_\mathrm{up}$\\
&  [ GHz ]  & [ K km s$^{-1}$ ] & [ km s$^{-1}$ ] & [ K ]\\
\\
\hline
\\
$J$=10--9 & 90.978 & 0.43(0.09) & 200 & 24\\
$J$=10--9 v$_7$=1$f$ & 91.333 & $<$0.12 & 170 & 342\\
$J$=12--11 & 109.173 & 0.75(0.16) & 300 & 34\\
$J$=12--11 v$_7$=1$e$ & 109.441 & 0.33(0.08) & 170 & 355\\
$J$=12--11 v$_7$=1$f$ & 109.598 & 0.21(0.08) & 170 & 355\\
\\
\hline
\end{tabular}}
\end{center}
\caption{\label{tab:ic860} Detected lines of HC$_3$N in IC~860. The integrated intensity is not corrected for main beam efficiency (see Table \ref{tab:bands}).}
\end{table}

\subsection{Models of galactic molecular emission}
\label{sec:models}

The line ratios observed by  \citet{baan08} were compared with a grid of XDR and PDR models by \citet{loenen08let}. These authors find that the low HNC/HCN$<$0.5 line ratios cannot be explained by XDR or PDR chemistry, and they claim that a possible explanation could come from mechanical heating. At temperatures $>$ 30 K, neutral-neutral reactions efficiently destroy HNC in favour of HCN \citep{hirota98}. \citet{loenen08let} suggest that shocks from supernova explosions may heat the gas to temperatures $>$100 K,  resulting in a depletion of HNC. The observed HNC/HCN line ratios can be explained by mechanical heating for a gas density of 10$^{4.5}$ and a star-formation rate (SFR) of 20 M$_\odot$ yr$^{-1}$. Higher densities would require higher SFR, which would result in the destruction of the molecular material surrounding the star-formation region \citep{loenen08let}. In this scenario, the observed low HNC/HCN line ratio would originate from low-density ($n_H<$10$^5$ cm$^{-3}$), warm (T$\simeq$100 K) gas. The corresponding high HCO$^+$/HCN observed in plot Fig. \ref{fig:plots}$b$ may be a consequence of the low gas density. As discussed in Sect. \ref{sec:hco+}, HCO$^+$ 1--0 has a lower critical density ($\simeq$10$^4$ cm$^{-3}$) compared with HCN 1--0 ($\simeq$10$^5$ cm$^{-3}$) and is more efficiently excited in diffuse gas. An increased HCO$^+$ abundance may also be caused by dissipation of turbulence driven by the supernovae winds. Models by \citet{joulain98} show that in low density gas, bursts of viscous dissipation can heat the gas up to the high temperatures required by endothermic reactions, such as C$^+$+H$_2\rightarrow$CH$^+$+H ($\Delta E/k\simeq$4600 K), more efficient than the standard cosmic ray-driven chemistry. This leads to an enhancement of CH$^+$, OH, and HCO$^+$, and may help to create the high HCO$^+$/HCN ratios we observe.\\
This picture is complicated by models of HCN and HNC chemistry as discussed in \citet{schilke92}. For gas temperatures $\simeq$100 K, an efficient destruction of HNC by neutral-neutral reactions requires densities higher than 10$^5$  cm$^{-3}$, while at lower densities the HCN/HNC abundance ratio is about unity \citep[Fig. 13 in ][]{schilke92}. However, at higher gas temperatures ($>$200 K), HNC destruction may become effective even for densities as low as $\simeq$10$^4$ cm$^{-3}$ and thus provide an explanation for the low HNC/HCN. 

\subsection{HCO$^+$, HNC, and PAHs}
\label{sec:hcopah}

In Fig. \ref{fig:plots}$d$, we compare the observed HNC/HCO$^+$ $J$=1--0 line ratios with the PAH equivalent widths (EW) from \citep{spoon07}. The distribution of HNC/HCO$^+$ can be described as a step function: it is roughly constant for low PAH EWs and has a sudden drop for Log$_\mathrm{10}$EW$>$-0.5. The galaxies with high PAH EWs are mostly starbursts or LIRGS. \\
PAHs are excited by UV radiation and destroyed by hard radiation (e.g., X-rays). Because if this,  strong PAH emission is usually interpreted as a clear signature of star-formation, instead of AGN activity. A decrement of HNC/HCO$^+$ for high PAH EWs may then imply either that  HCO$^+$ is enhanced in a PDR, or that HNC is depleted. Indeed, if we compare the HCO$^+$ and HNC emission to HCN 1--0, for increasing PAH EWs we have an increase in the HCO$^+$/HCN ratio and a decrease in HNC/HCN. This can be seen in the graphs in Fig. \ref{fig:hncdep}.\\
If directly connected to the molecular abundances, the observed ratios are difficult to explain with PDR and XDR chemistry. Models by \citet{meijerink05} show that  HCO$^+$/HCN is lower in PDRs than in XDRs, and HNC/HCN very close to unity in a wide range of densities and gas temperatures. However, the observed trend agrees with results by \citet{imanishi07}, who find that AGNs have a lower HCO$^+$/HCN ratio than starburst galaxies. \\ A depletion of HNC is also expected in shocks. Here, because of the high temperatures and densities, neutral-neutral reactions efficiently destroy HNC in favour of HCN  \citep{schilke92}. Mechanical heating from shocks generated by supernova explosions may explain the observed trend. In this case, the HNC/HCN ratio should follow the dependence on temperature of neutral-neutral reactions. In Fig. \ref{fig:plots}$e$ we compare the HNC/HCN ratio with the ratio of the IRAS fluxes at 60 and 100 $\mu$m. The graph clearly shows that there is no strong dependence of the HNC/HCN ratio with dust temperature. At densities $>$10$^5$ cm$^{-3}$, gas and dust thermalise and this should show in the diagram, since neutral-neutral reactions strongly depend on gas temperature. At densities below 10$^5$ cm$^{-3}$, the thermal coupling becomes poor, however, and it is possible to have T$_\mathrm{gas}>$T$_\mathrm{dust}$. This scenario is complicated by observations of HCO$^+$ in shocked gas in the Galaxy \citep{hcoschock}, which reveal that the molecule is destroyed by reactions with H$_2$O in the shock front. Detectable HCO$^+$ abundances are found in the aftermath of the shock, where water freezes on grains. \citet{hcoschock} report an HCO$^+$/HCN column density ratio of 0.6 in the outflow of the low-mass protostar NGC~1333~IRAS~2A. 

\section{Conclusions}
\label{sec:conclusions}

We observed the molecular emission of 23 galaxies with the new EMIR receiver at IRAM 30m in the frequency bands 88 GHz and 112 GHz. The observed line ratios were compared with existing models of emission from PDRs, XDRs, and hot cores. 
All non-compact starburst galaxies are found to be HCO$^+$-luminous, with line ratios HCO$^+$/HCN $\geq$ 1. Luminous infrared galaxies and Seyferts  do not show any strong trend. However, the majority of active galactic nuclei do reside at low HCO$^+$/HCN values compared with starbursts. \\
When HNC/HCO$^+$ increases, galaxies move towards higher HNC/HCN and HCO$^+$/HCN ratios. Galaxies with HCO$^+$/HCN $<1$ present on average a 30 \% increase in HNC/HCN, compared with those at higher HCO$^+$/HCN ratios. If we exclude the Seyfert NGC~2273, the brightest  HNC emission is found in luminous infrared galaxies. Starbursts are generally characterized by faint HNC and bright HCO$^+$ emission (compared with HCN), while LIRGs mostly occupy the opposite end of the HNC-HCO$^+$ correlation, with high HNC and low HCO$^+$ intensities. All HNC-bright galaxies are either LIRGs or Seyferts. \\
All HC$_3$N detections have HCO$^+$/HCN $<1$ and HNC/HCN$>$0.4. Most of these are are LIRGs, the only exception being the Seyfert NGC~1068.\\
Starburst galaxies tend to have high HCO$^+$/HCN and high PAH EWs, in agreement with observations by \citet{kohno2001} and \citet{imanishi07}.\\
The interpretation of the low HCO$^+$/HCN ratio as an effect of XDR chemistry \citep{imanishi07} is inconsistent with our observations, since we find bright HC$_3$N emission in galaxies with the lowest HCO$^+$/HCN. The HC$_3$N molecule is easily destroyed by reactions with ions and hard UV and is thus extremely unlikely to survive in X-ray dominated environments.\\
 Models by \citet{bayet08} for hot cores provide a better explanation for the observed ratios, suggesting that bright HC$_3$N and faint HCO$^+$ emission might be good tracers of embedded star-formation.\\
Vibrationally excited HC$_3$N was detected in IC~860, where it was possible to obtain an estimate of the rotational temperature of the $v$=0 transitions of 15 K, and an upper limit for the $v_7$=1 levels of 42 K. This galaxy has the highest HC$_3$N/HCN ratio of the sample, representing an interesting target for further investigation of the molecule's properties.\\
In agreement with previous studies, we find that the CO/$^{13}$CO intensity ratio is positively correlated with dust temperature, with typical values $\simeq$ 20, higher than the ones observed in the Galaxy. We suggest that these high values are caused by either high gas temperatures or infall of unprocessed material in the centres of starbursts.\\
The emission of CN and C$^{18}$O appear to be positively correlated. This may simply be because both CN and C$^{18}$O are optically thin and scale with the total gas column, or may be caused by ISM enrichment by the starburst activity. CN emission is, indeed, particularly strong in PDR surfaces, which are supposed to be tracing the impact of star-formation on the host clouds.  \\
In agreement with previous extragalactic studies \citep[e.g., ][]{baan08}, we find HNC/HCN intensity ratios lower than 0.4, which cannot be explained by current models of PDR and XDR chemistry. A possible explanation was suggested by \citet{loenen08let} to be mechanical heating by supernova explosions. This model would require low gas densities ($<$10$^5$ cm$^{-3}$) or extremely high star-formation rates ($>$20 M$_{\odot}$), but would explain both the low HNC/HCN and high HCO$^+$/HCN ratios in starbursts. Mechanical heating by SN explosions could also explain the observed drop of the HNC/HCO$^+$ ratio for PAH EWs $>$ 0.3 $\mu m$, if we assume that PAHs efficiently trace starformation.\\
However, a temperature dependence of the HNC/HCN ratio, which would be expected if the HNC abundance was depleted by neutral-neutral reaction in the shocked gas, does not emerge from our data. Furthermore, it is not yet clear how SN shocks affect HCO$^+$ abundance, with different models giving contrasting results.

\subsection{Future directions}

Evidently, explaining the molecular emission observed in galaxies by means of the existing chemical models is a challenging task.
Our understanding of the gas physical conditions is still affected by many uncertainties. Mechanical heating, invoked to explain the low HNC/HCN ratios, requires further investigation with shock models  and observations of shock tracers (e.g., IR transitions of H$_2$ or SiO). For single dish observations, the analysis of the observed emission is complicated by the lack of spatial information. Radiation from different molecules and transitions may in fact be coming from different regions in the galaxy, and thus  it may be  impossible to  explain observations with only one set of physical conditions. Spatial distribution and density gradients in the molecular gas may have a strong impact on the observed line ratios. This may be particularly important for the  HCO$^+$/HCN ratio, which appears to be strongly density-dependent, because of both chemical and excitation effects. The angular resolution needed to resolve the spatial distribution of molecular gas in distant galaxies will become available in the foreseeable future with the ALMA interferometer. Observations of mm lines of formaldehyde (H$_2$CO) may also help to constrain density and temperature in these objects. Sub-mm observations of high-J transitions of CO, which have been indicated as excellent tracers of PDRs by \citet{meijerink07}, are now available with the Herschel telescope. The increased sensitivity of new generation mm- and sub-mm observatories will allow us to detect less abundant molecular species (e.g., HNCO, CH$_3$OH) that may be better tracers of interstellar chemistry \citep[e.g., ][]{meier05,martin09}. The evolution of the starburst phase can be studied with radio continuum observations, which will be discussed in Paper II. We also plan to further explore molecular excitation in our sample by means of multi-transition studies. We successfully applied for telescope time at IRAM 30m to observe 1 mm lines of the species discussed in the present work. Collected data will be discussed in Paper III, where  we will present a NLTE analysis of the emission of the main molecular species, together with a comparison with a grid of PDR and XDR models. 

\begin{acknowledgements}

We thank the staff at the IRAM 30m telescope for their kind help and
support during our observations. Furthermore, we would like to thank the 
IRAM PC for their generous allocation of time for this project. 

This research was supported by the EU Framework 6 Marie Curie Early Stage Training programme under contract number MEST-CT-2005-19669 "ESTRELA" and by the European Community Framework Programme 7, Advanced Radio Astronomy in Europe, grant agreement no. 227290, "RadioNet".
AA, MAPT and MR acknowledge support by the Spanish MICINN through grant AYA2009-13036-CO2-01.
PvdW would like to thank the Scottish Universities Physics Alliance (SUPA) for support through a Distinguished Visitor Grant.

\end{acknowledgements}

\bibliographystyle{aa} 
\bibliography{bibliototale}

\begin{appendix}
\section{Detection of HC$_3$N}
\label{sec:dethc3n}
\begin{figure}
\begin{centering}
\includegraphics[width=.45\textwidth,keepaspectratio]{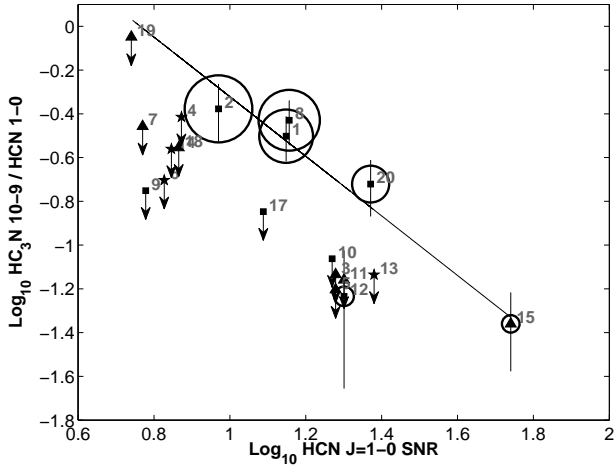}
\caption{\label{fig:hc3n} Ratio of HC$_3$N $J$=10--9 over HCN $J$=1--0 integrated intensities against the HCN $J$=1--0 signal-to-noise ratio. The line shows a linear fit through the HC$_3$N detections. See Fig. \ref{fig:plots} for an explanation of the symbols.}
\end{centering}
\end{figure}

The ratio of HC$_3$N $J$=10--9 over HCN $J$=1--0 integrated intensities is plotted in Fig. \ref{fig:hc3n} against the HCN $J$=1--0 signal to noise ratio (SNR). The graph clearly shows that the non--detections are not caused by the sensitivity limit of our observations. In a wide range of SNR values, detections and non--detections coexist. \\ The line in Fig. \ref{fig:hc3n} shows a first order fit through the HC$_3$N detections, excluding the low--SNR detection in NGC~7771 (number 12 on the graph). Our detection limits (arrows in the graph) are, on average, 3 times fainter than the expected intensity derived by the fit. This means that in most of the cases -- this may not be true for NGC~2273, number 19 in the graph -- we do not detect HC$_3$N not because of a lack of sensitivity, but because the emission is intrinsically faint. In NGC~1068 the molecular emission is about one order of magnitude brighter than in the other galaxies of the sample, therefore the HC$_3$N detection (with the lowest HC$_3$N/HCN ratio in the sample) in this galaxy may be only caused by its intrinsic high luminosity.
\vfill

\section{HCO$^+$, HNC and PAHs}
\label{sec:hncdep}

\begin{figure}
\begin{centering}
\includegraphics[width=.45\textwidth,keepaspectratio]{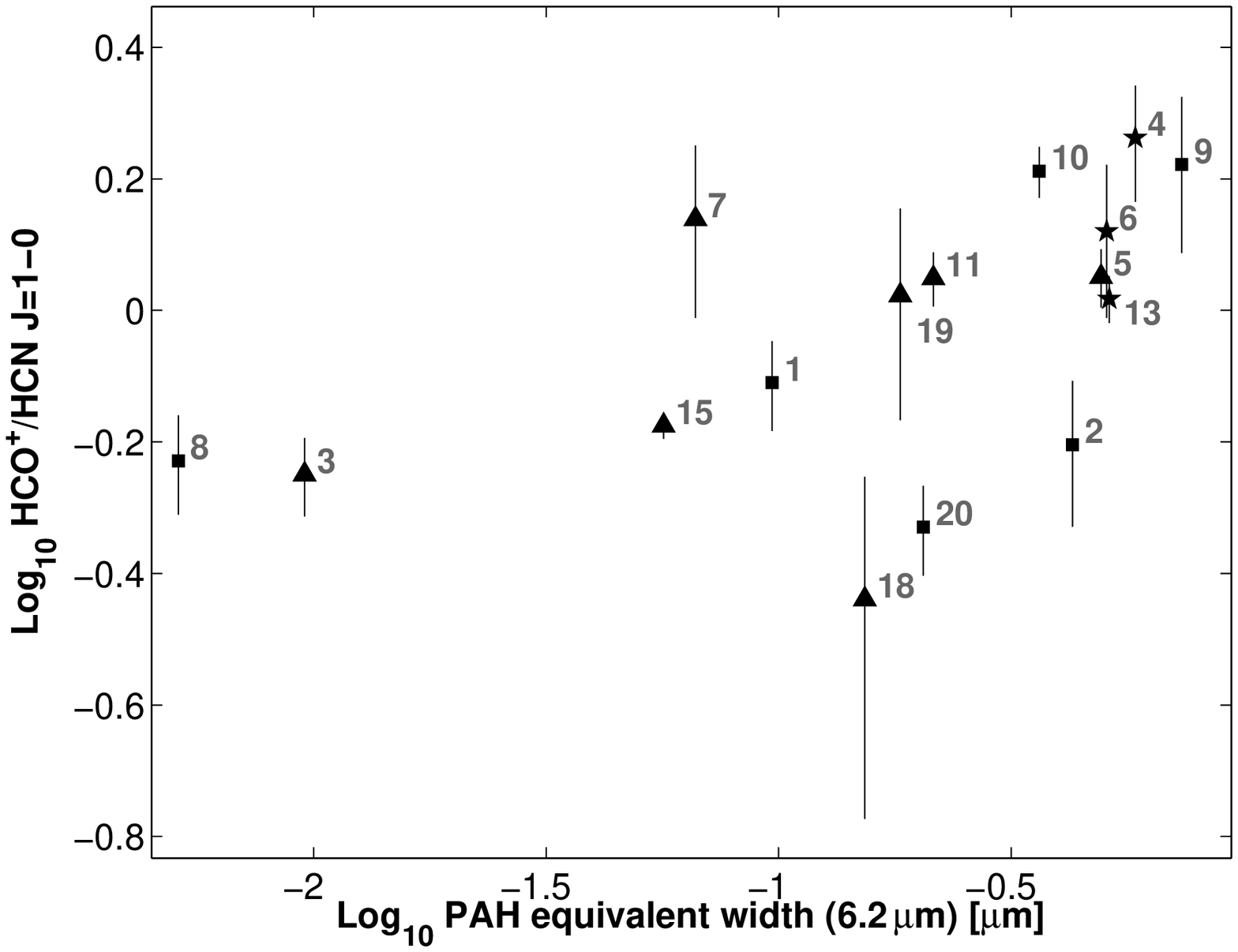}
\includegraphics[width=.45\textwidth,keepaspectratio]{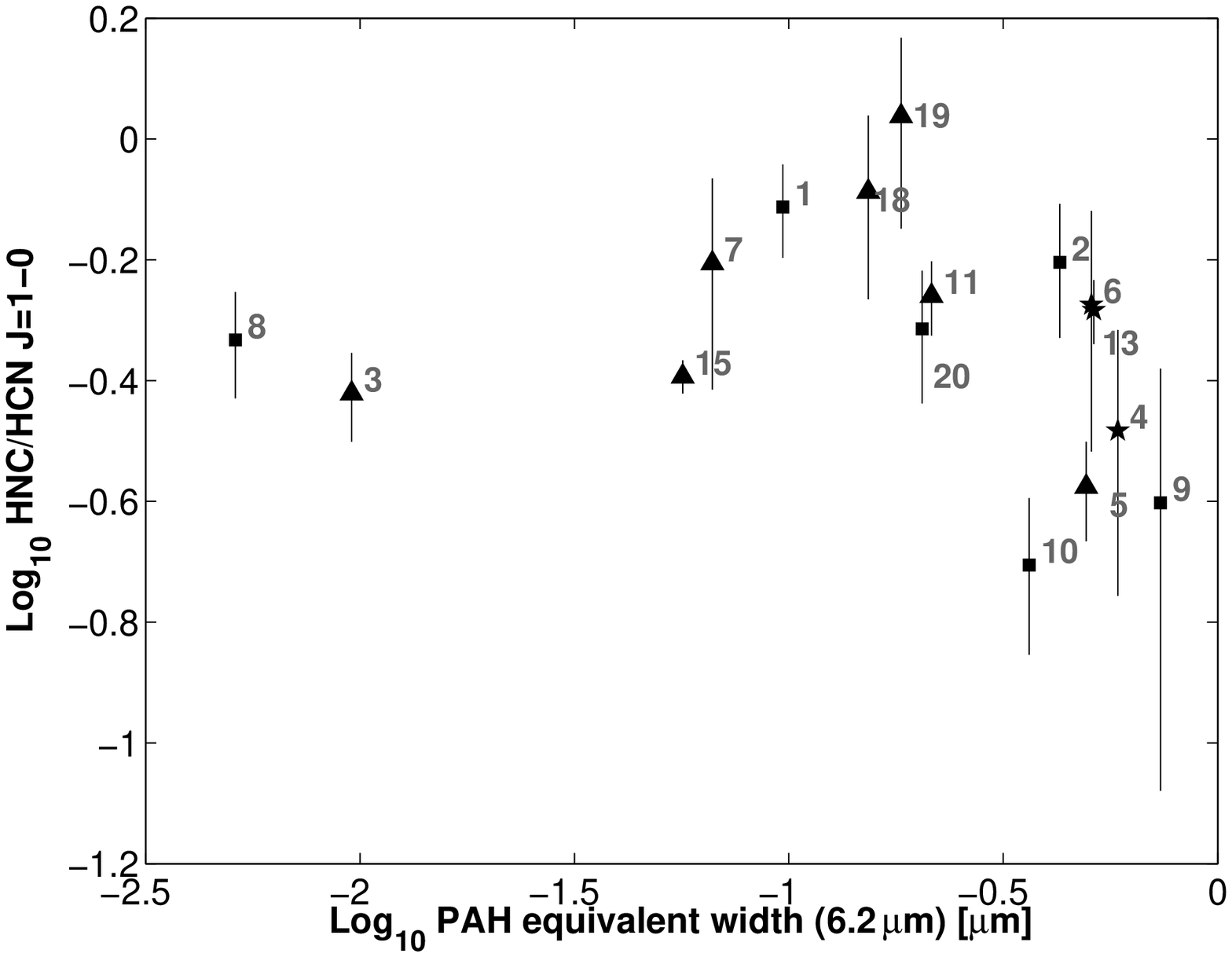}
\caption{\label{fig:hncdep} Line ratios of HCO$^+$ and HNC over HCN against the PAH equivalent width. Notice the decrease of HNC and increase of HCO$^+$ at high PAH EWs, to be compared with Fig. \ref{fig:plots}$d$.}
\end{centering}
\end{figure}

The strong dependence of the HCO$^+$/HNC ratio with PAH equivalent width, shown in Fig. \ref{fig:plots}$d$, can be better understood by plotting the behaviour of the single tracers with increasing PAH EW. \\ In Fig. \ref{fig:hncdep} we plot the ratios HCO$^+$/HCN and HNC/HCN against the PAH strength. Evidently, at high PAH EWs we have a depletion of HNC (respect to HCN) and an enhancement of HCO$^+$. This two trends both contribute to the fast decrease of the HCO$^+$/HNC ratio seen in Fig. \ref{fig:plots}$d$. For a discussion, see text in Section \ref{sec:hcopah}.

\section{Comparison with the literature}
\label{sec:comptab}
In Table  \ref{tab:comp} we compare our observations with literature data from \citet{baan08}. Antenna temperatures, obtained with EMIR, were translated into flux densities, by applying the Jy/K conversion factor
\begin{equation}
J[Jy]=T_\mathrm{A}^\star[K]\times 3.906 \times F_\mathrm{eff}/\eta_\mathrm{A},
\end{equation}  
with $F_\mathrm{eff}$ forward efficiency, and $\eta_\mathrm{A}$ antenna efficiency of the 30m telescope at 88 GHz.

\begin{table*}
\renewcommand{\arraystretch}{1.2}
\begin{center}
\begin{tabular}{l c c| c c |c c |c c c| c c c } 
\hline
\hline
Galaxy &\multicolumn{2}{c}{HCN 1-0} & \multicolumn{2}{c}{HNC 1-0} & \multicolumn{2}{c}{HCO$^+$ 1-0} & \multicolumn{3}{c}{HNC/HCN} & \multicolumn{3}{c}{HCO$^+$/HCN} \\ 
 & \multicolumn{2}{c}{[Jy km s$^{-1}$]} & \multicolumn{2}{c}{[Jy km s$^{-1}$]} & \multicolumn{2}{c}{[Jy km s$^{-1}$]} \\ 
& B08 & EMIR & B08 & EMIR & B08 & EMIR & B08 & EMIR & $\Delta\%$ & B08 & EMIR & $\Delta\%$ \\ 
 NGC1068 & 158 & 130 & 61.9 & 52.4 & 143 & 86.6 & 0.39 & 0.40 & 3 & 0.91 & 0.67 & 26 \\ 
 NGC660 &  33 & 28.3 & 18.9 & 14.7 & 32.8 & 29.5 & 0.57 & 0.52 & 9 & 0.99 & 1.04 & 5 \\ 
 Mrk231 & 17.6 & 11.2 & 12.1 & 4.24 & 12.6 & 6.3 & 0.69 & 0.38 & 45 & 0.72 & 0.56 & 21 \\ 
 NGC6240 & 31.2 & 16.4 &   8 & 3.24 & 23.4 & 26.7 & 0.26 & 0.2 & 23 & 0.75 & 1.63 & 117 \\ 
 NGC1614 & 7.2 & 4.83 &   - & 1.59 &   - & 8.84 &   - & 0.33 &   - &   - & 1.83 &   - \\ 
 UGC5101 & 10.4 & 6.48 &   - & 5.3 &   - & 2.36 &   - & 0.82 &   - &   - & 0.36 &   - \\ 
 NGC3079 & 43.1 & 24.6 & 32.3 & 6.54 &   - & 27.7 & 0.75 & 0.27 & 65 &   - & 1.12 &   - \\ 
 NGC3556 & 16.8 & 3.18 &   - & 1.18 &   - & 5.01 &   - & 0.37 &   - &   - & 1.57 &   - \\ 
 NGC7469 & 10.8 & 11.8 &  17 & 6.48 & 15.8 & 13.2 & 1.57 & 0.55 & 65 & 1.46 & 1.12 & 23\\ 
 NGC7771 & 22.7 & 21.2 &   - & 9.42 & 21.5 &  20 &   - & 0.44 &   - & 0.95 & 0.94 & 0.3 \\ 
\\
\hline
\end{tabular} 
\end{center}

\caption{\label{tab:comp} Comparison with data from the literature, as reported by \citet{baan08} (B08), and this work (EMIR). Exact references for the observations can be found in Table B1 in  \citet{baan08}. Differences (in \%) between B08 and EMIR line intensity ratios are shown in column $\Delta\%$.}
\end{table*}

\section{EMIR Spectra}
\label{sec:spectra}

\begin{figure*}[h][h]
\centering
\includegraphics[width=0.9\textwidth,keepaspectratio]{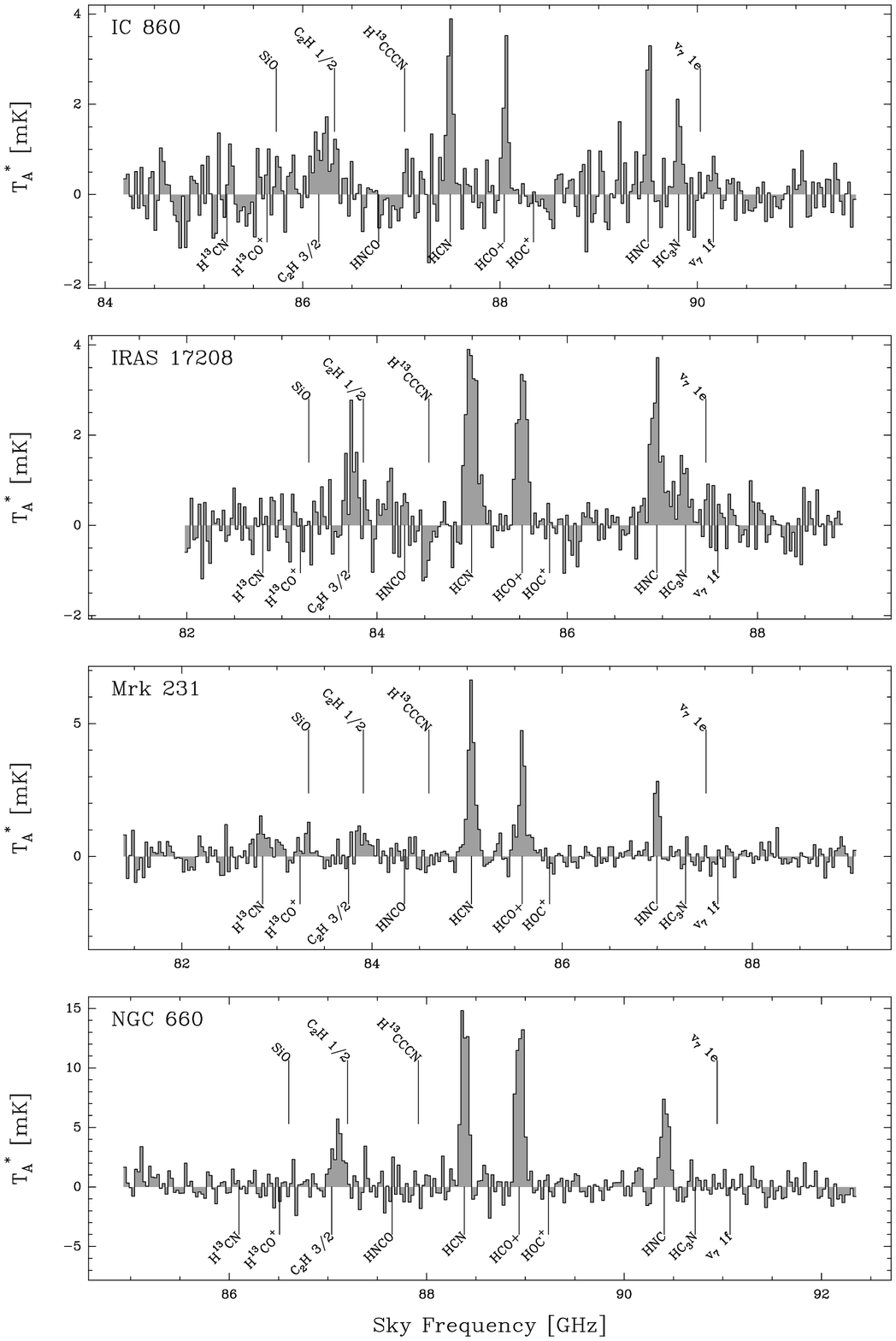}
\caption{\label{fig:88ghz}Observed spectra in the 88 GHz band. The intensity scale is in T$_A^\star$, not corrected for main beam efficiency. The main molecular  transitions are labelled regardless of line detection. The C$_2$H 3/2 and 1/2 labels mark the limits of the C$_2$H multiplet at 87 GHz. Transitions of vibrationally excited HC$_3$N are labelled as $v_71e$ and $v_71f$. The frequency scale is the observed frequency, not corrected for redshift.}
\end{figure*}

In Fig. \ref{fig:88ghz}--\ref{fig:112last} we show the spectra observed in all the galaxies in the sample. The intensities are in T$_A^\star$, not corrected for beam efficiency (see Table \ref{tab:bands}). On the $x$-axis we plot the sky frequency in GHz, not corrected for redshift. The position of the main emission lines is also shown. 

\begin{figure*}[h]
\centering
\includegraphics[width=0.9\textwidth,keepaspectratio]{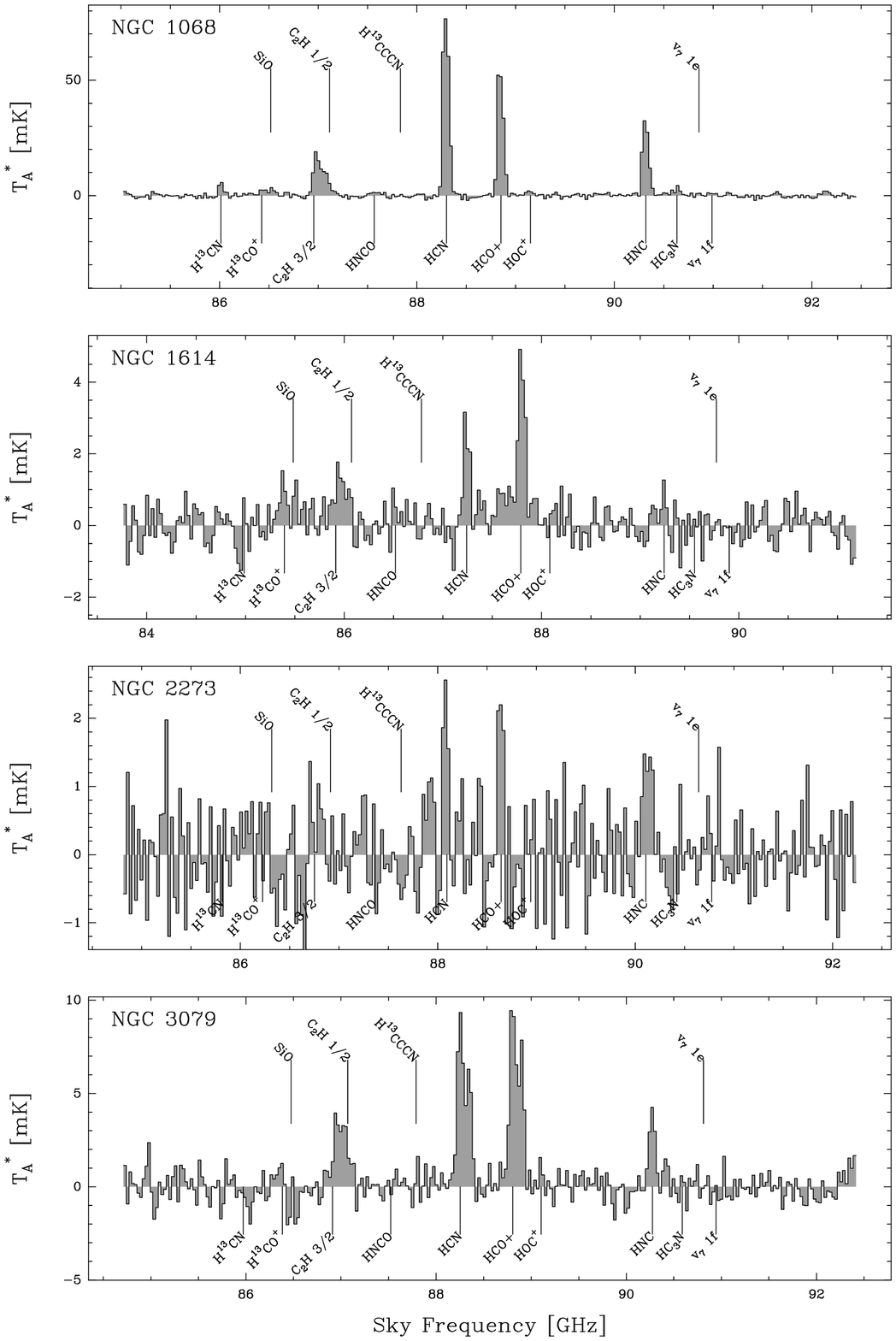}
\caption{Observed spectra in the 88 GHz band ($continued$). See caption of Fig. \ref{fig:88ghz}.}
\end{figure*}
\begin{figure*}[h]
\centering
\includegraphics[width=0.9\textwidth,keepaspectratio]{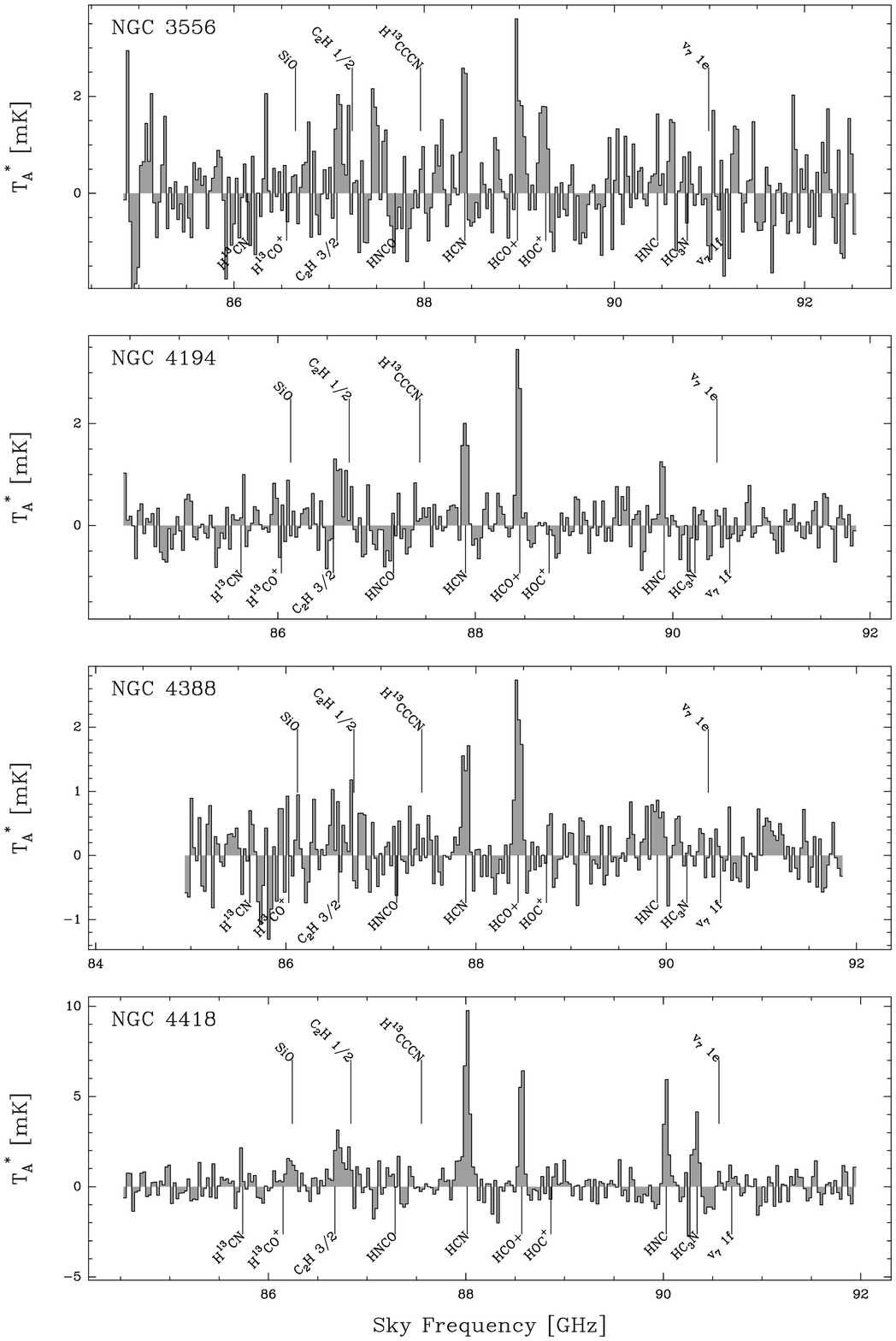}
\caption{Observed spectra in the 88 GHz band ($continued$). See caption of Fig. \ref{fig:88ghz}.}
\end{figure*}
\begin{figure*}[h]
\centering
\includegraphics[width=0.9\textwidth,keepaspectratio]{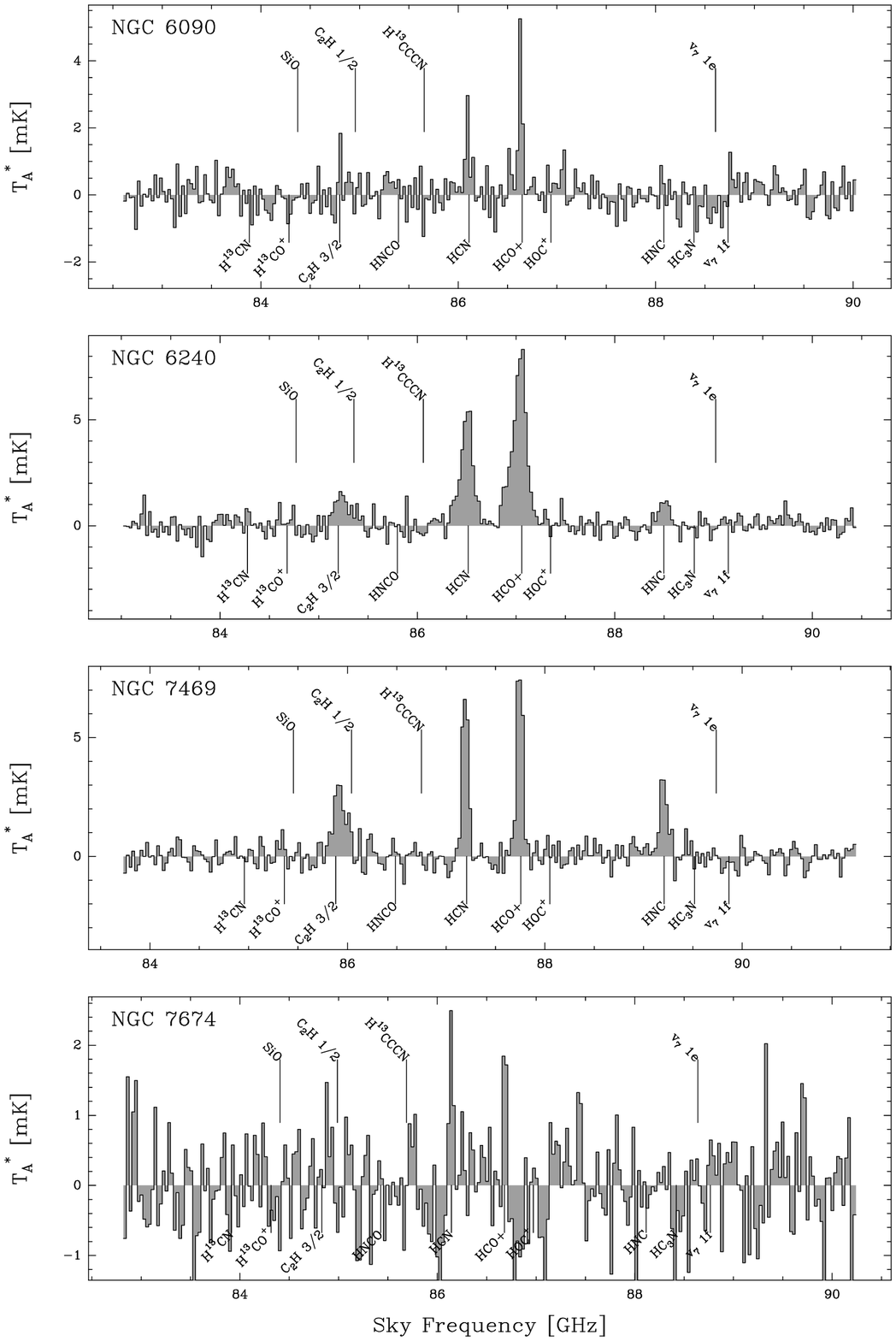}
\caption{Observed spectra in the 88 GHz band ($continued$). See caption of Fig. \ref{fig:88ghz}.}
\end{figure*}
\begin{figure*}[h]
\centering
\includegraphics[width=0.9\textwidth,keepaspectratio]{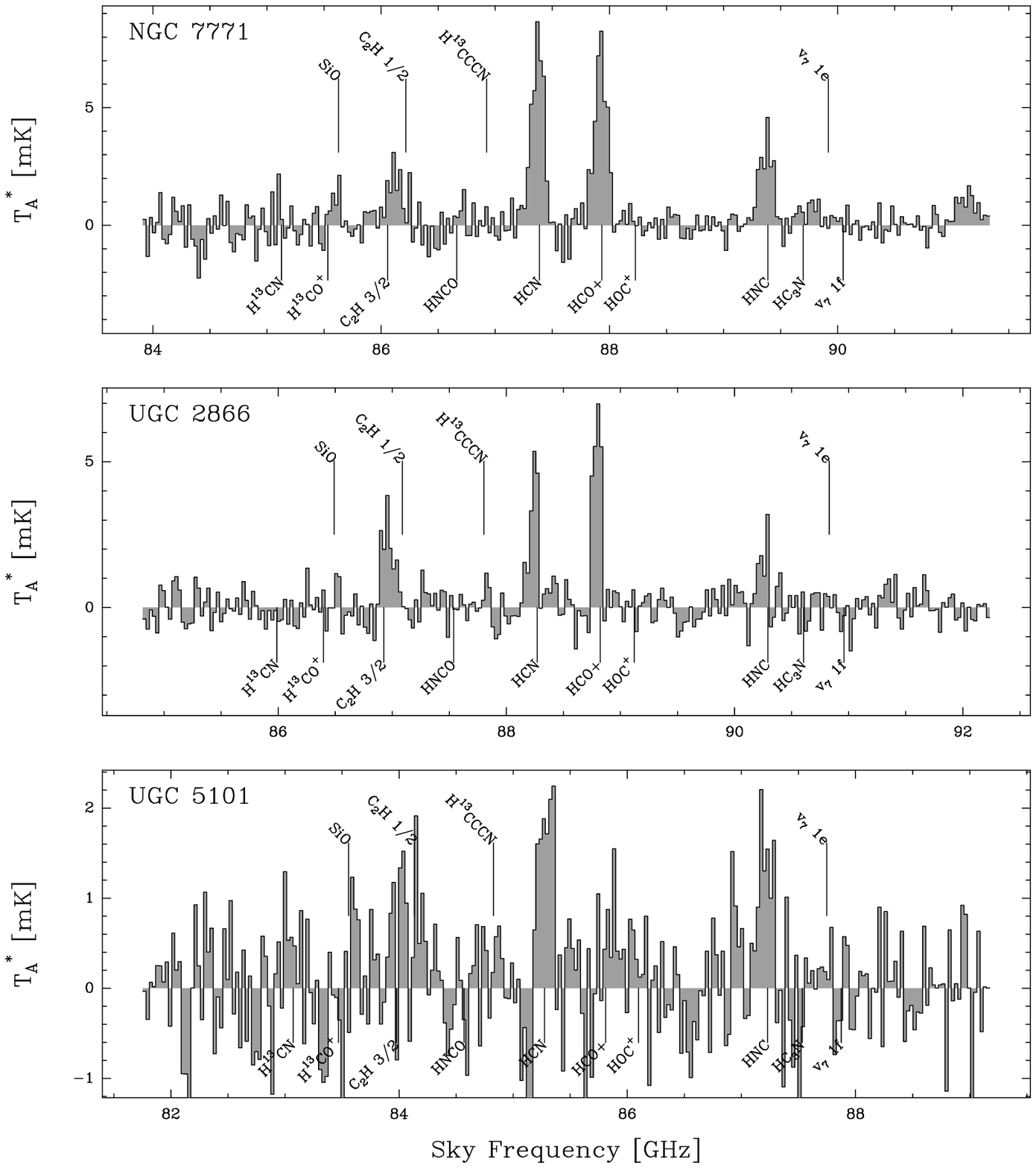}
\caption{Observed spectra in the 88 GHz band ($continued$). See caption of Fig. \ref{fig:88ghz}.}
\end{figure*}
\begin{figure*}[h]
\centering
\includegraphics[width=0.9\textwidth,keepaspectratio]{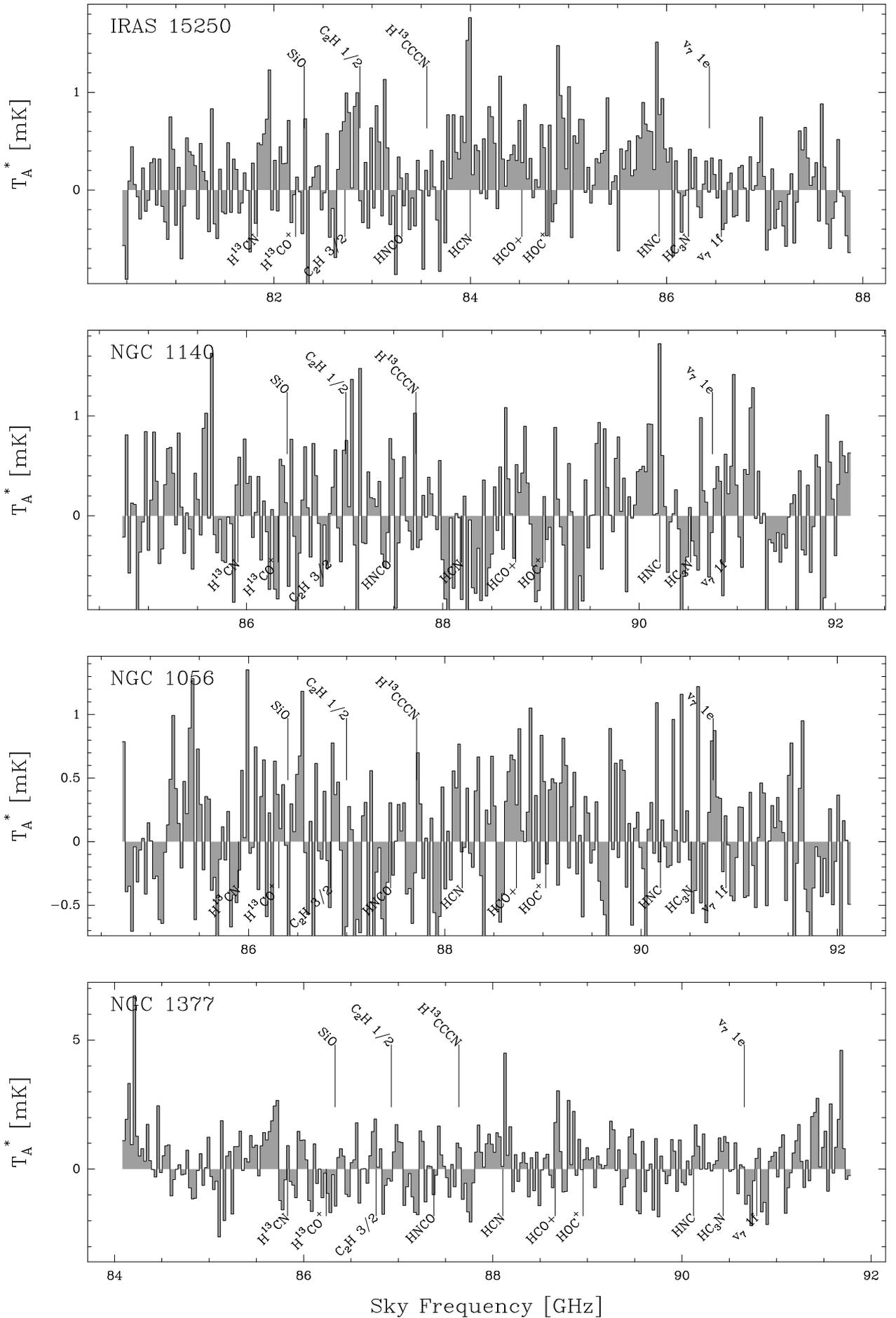}
\caption{Non-detections in the 88 GHz band. See caption of Fig. \ref{fig:88ghz}.}
\end{figure*}

\renewcommand{\arraystretch}{.9}
\begin{figure*}[h]
\centering
\includegraphics[width=0.9\textwidth,keepaspectratio]{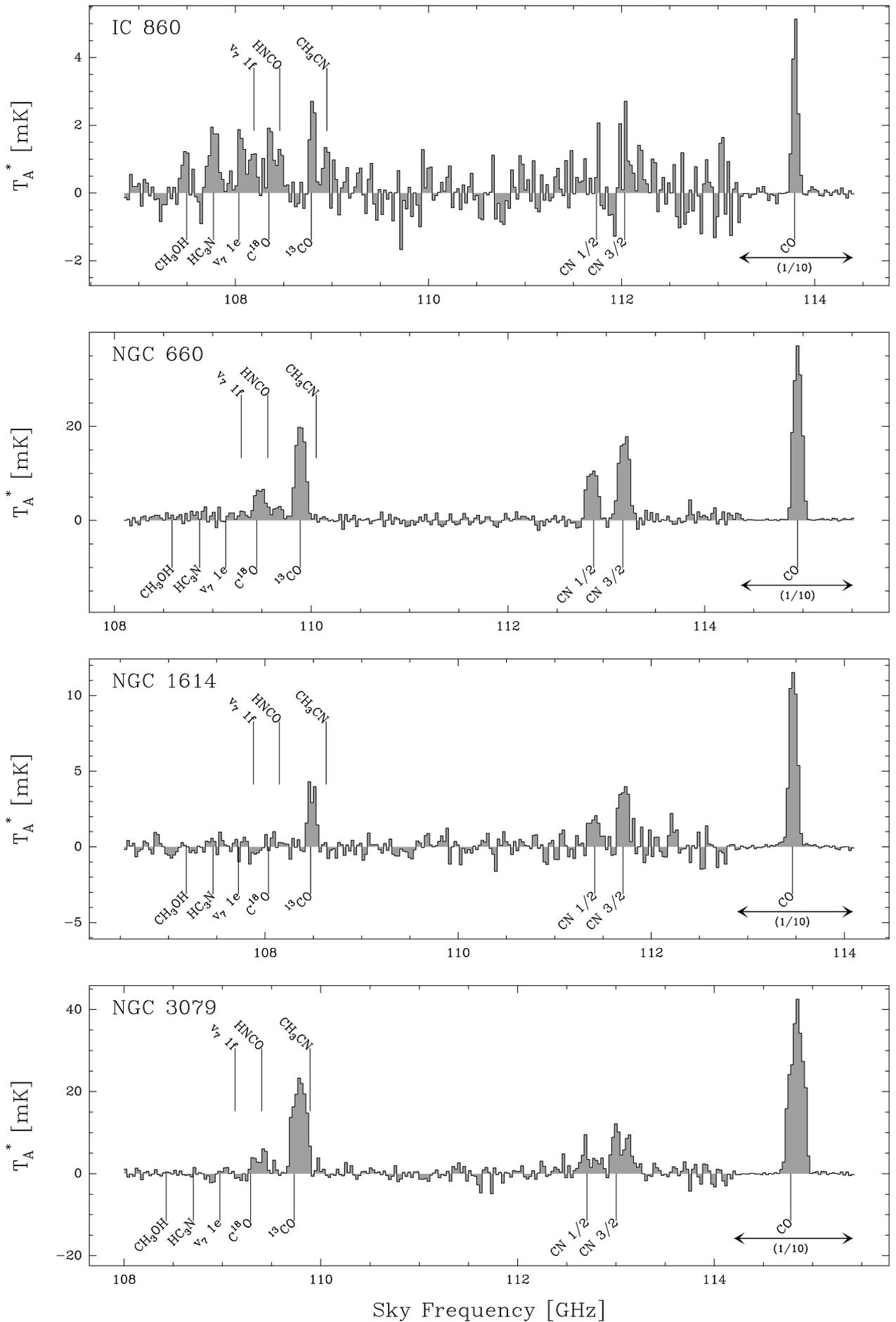}
\caption{\label{fig:112ghz}Observed spectra in the 112 GHz band. The intensity scale is in T$_A^\star$, not corrected for main beam efficiency. The region marked with $(1/10)$ around the CO 1-0 line was scaled down by a factor 10. The main molecular  transitions are labelled regardless of line detection. Transitions of vibrationally excited HC$_3$N are labelled as $v_71e$ and $v_71f$. The frequency scale is the observed frequency, not corrected for redshift.}
\end{figure*}
\begin{figure*}[h]
\centering
\includegraphics[width=0.9\textwidth,keepaspectratio]{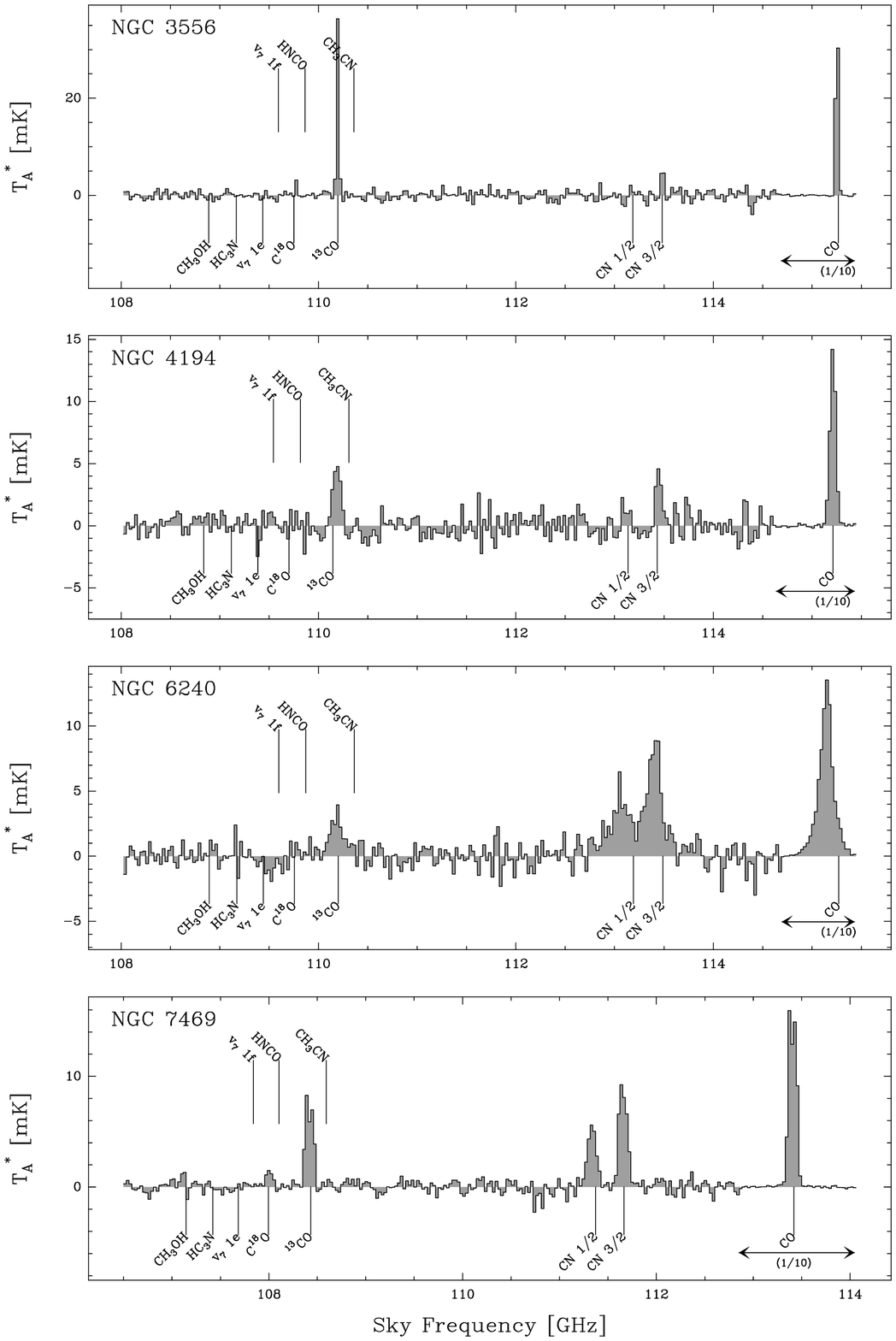}
\caption{Observed spectra in the 112 GHz band ($continued$). See caption of Fig. \ref{fig:112ghz}.}
\end{figure*}
\begin{figure*}[h]
\centering
\includegraphics[width=0.9\textwidth,keepaspectratio]{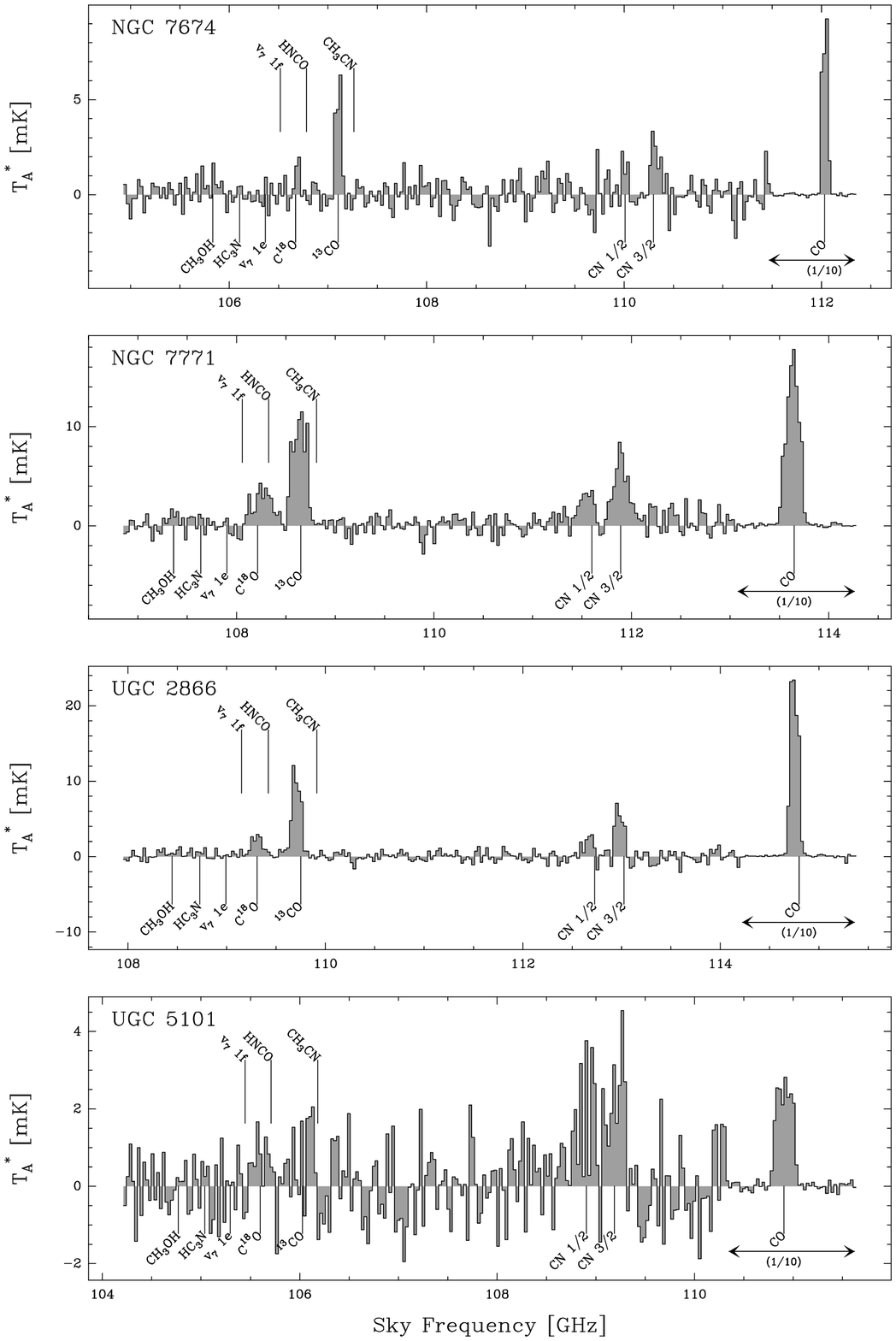}
\caption{\label{fig:112last} Observed spectra in the 112 GHz band ($continued$). See caption of Fig. \ref{fig:112ghz}.}
\end{figure*}
\newpage
\section{Derived line properties}
\label{sec:tables}
\begin{table*}[h]
\centering
\begin{tabular}{l l c c | l c c} 
\hline 
\hline 
& & & & & & \\ 
 Galaxy & Transition & $\int T^{\star}_A dv$ & $\Delta_V$ & Transition & $\int T^{\star}_A dv$ & $\Delta_V$ \\ 
 & & K km s$^{-1}$ & km s$^{-1}$ & & K km s$^{-1}$ & s$^{-1}$ \\  
& & & & & & \\ 
\hline 
& & & & & & \\ 
& 88 GHz Band & & & 112 GHz Band & & \\ 
& & & & & & \\ 
I17208 & H13CO+1-0 &  $<$2.55 & - & & &  \\
 & SiO2-1 &  $<$0.23 & - & & &  \\
 & H13CN1-0 &  $<$0.23 & - & & &  \\
 & C2H1-0 &  0.82(0.14) & 388(80)$^\mathrm{\star}$ & & &  \\
 & HCN1-0 &  1.97(0.14) & 452(35) & & &  \\
 & HCO+1-0 &  1.53(0.13) & 407(38) & & &  \\
 & HNC1-0 &  1.52(0.16) & 473(65) & & &  \\
 & HC3N10-9 &  0.62(0.12) & 452(70) & & &  \\
& & & & & & \\ 
IC860 & H13CO+1-0 &  $<$0.20 & - &  CH3OH &  0.40(0.13) & 312(104) \\
 & SiO2-1 &  $<$0.20 & - &  HC3N12-11 &  0.75(0.16) & 300(82) \\
 & H13CN1-0 &  $<$0.20 & - &  HC3N12-11v7e &  0.33(0.08) & 170(60) \\
 & & & & HC3N12-11v7f & 0.21(0.08) & 170(60) \\
 & C2H1-0 &  1.86(0.36) & 1375(416)$^\mathrm{\star}$ &  C18O1-0 &  0.50(0.20) & 253(150) \\
 & HCN1-0 &  1.12(0.12) & 239(31) &  13CO1-0 &  0.54(0.10) & 162(32) \\
 & HCO+1-0 &  0.70(0.10) & 202(38) &  CN1-0J=1/2 &  $<$0.19 & - \\
 & HNC1-0 &  0.60(0.10) & 171(29) &  CN1-0J=3/2 &  $<$0.27 & - \\
 & HC3N10-9 &  0.43(0.09) & 200(67) &  CO1-0 &  10.24(0.17) & 179(3) \\
& & & & & & \\ 
Mrk231 & H13CO+1-0 &  0.14(0.1) & 271(62) & & &  \\
 & SiO2-1 &  0.25(0.11) & 187(62) & & &  \\
 & H13CN1-0 &  0.34(0.1) & 228(68) & & &  \\
 & C2H1-0 &  0.38(0.20) & 513(304)$^\mathrm{\star}$ & & &  \\
 & HCN1-0 &  1.90(0.10) & 288(21) & & &  \\
 & HCO+1-0 &  1.07(0.09) & 208(23) & & &  \\
 & HNC1-0 &  0.72(0.08) & 203(25) & & &  \\
 & HC3N10-9 &  $<$0.14 & - & & &  \\
& & & & & & \\ 
NGC1614 & H13CO+1-0 &  0.36(0.13) & 264(62) &  CH3OH &  $<$0.14 & - \\
 & SiO2-1 &  0.24(0.13) & 270(62) &  HC3N12-11 &  0.14(0.14) & 264(237) \\
 & H13CN1-0 &  $<$0.28 & - &  HC3N12-11v7e &  $<$0.14 & - \\
 & C2H1-0 &  0.28(0.35) & 264(270)$^\mathrm{\star}$ &  C18O1-0 &  $<$0.14 & - \\
 & HCN1-0 &  0.79(0.29) & 261(31) &  13CO1-0 &  1.12(0.12) & 254(26) \\
 & HCO+1-0 &  1.50(0.30) & 290(75) &  CN1-0J=1/2 &  0.62(0.13) & 305(58) \\
 & HNC1-0 &  0.24(0.23) & 162(200) &  CN1-0J=3/2 &  1.31(0.14) & 309(33) \\
 & HC3N10-9 &  $<$0.28 & - &  CO1-0 &  33.71(0.15) & 267(2) \\
& & & & & & \\ 
NGC3079 & H13CO+1-0 &  0.33(0.14) & 250(40) &  CH3OH &  $<$1.38 & - \\
 & SiO2-1 &  $<$0.28 & - &  HC3N12-11 &  $<$1.38 & - \\
 & H13CN1-0 &  $<$0.28 & - &  HC3N12-11v7e &  $<$1.38 & - \\
 & C2H1-0 &  2.26(0.24) & 575(68)$^\mathrm{\star}$ &  C18O1-0 &  2.00(0.20) & 254(16) \\
 & HCN1-0 &  4.18(0.22) & 499(30) &  13CO1-0 &  10.70(0.20) & 350(16) \\
 & HCO+1-0 &  4.70(0.23) & 516(27) &  CN1-0J=1/2 &  0.88(0.40) & 85(16) \\
 & HNC1-0 &  1.11(0.15) & 230(27) &  CN1-0J=3/2 &  4.40(0.60) & 200(16) \\
 & HC3N10-9 &  $<$0.28 & - &  CO1-0 &  183.00(9.00) & 200(16) \\
& & & & & & \\ 
NGC4194 & H13CO+1-0 &  $<$0.09 & - &  CH3OH &  $<$0.09 & - \\
 & SiO2-1 &  $<$0.09 & - &  HC3N12-11 &  $<$0.09 & - \\
 & H13CN1-0 &  $<$0.09 & - &  HC3N12-11v7e &  $<$0.09 & - \\
 & C2H1-0 &  0.45(0.10) & 415(97)$^\mathrm{\star}$ &  C18O1-0 &  0.09(0.20) & 220(30) \\
 & HCN1-0 &  0.47(0.07) & 177(25) &  13CO1-0 &  1.50(0.14) & 243(30) \\
 & HCO+1-0 &  0.62(0.07) & 151(19) &  CN1-0J=1/2 &  0.50(0.14) & 200(30) \\
 & HNC1-0 &  0.25(0.07) & 175(52) &  CN1-0J=3/2 &  1.18(0.15) & 219(30) \\
 & HC3N10-9 &  $<$0.09 & - &  CO1-0 &  28.65(0.17) & 180(30) \\
& & & & & & \\ 
\end{tabular}

\caption{\label{tab:linprop} Summary of the line properties of the observed galaxies. Integrated intensities and line widths were derived by means of Gaussian fitting. Uncertainties are shown in parenthesis. $(\mathrm{\star})$:  For C$_2$H, the integrated intensity is obtained by fitting a single Gaussian to the multiplet at 87 GHz. This results in large line widths, which are not to be attributed to dynamics.}
\end{table*}
\begin{table*}
\centering
\begin{tabular}{l l c c | l c c} 
\hline 
\hline 
& & & & & & \\ 
 Galaxy & Transition & $\int T^{\star}_A dv$ & $\Delta_V$ & Transition & $\int T^{\star}_A dv$ & $\Delta_V$ \\ 
 & & K km s$^{-1}$ & km s$^{-1}$ & & K km s$^{-1}$ & s$^{-1}$ \\  
& & & & & & \\ 
\hline 
& & & & & & \\ 
& 88 GHz Band & & & 112 GHz Band & & \\ 
& & & & & & \\ 
NGC4388 & H13CO+1-0 &  $<$0.10 & - & & &  \\
 & SiO2-1 &  $<$0.10 & - & & &  \\
 & H13CN1-0 &  $<$0.10 & - & & &  \\
 & C2H1-0 &  0.10(0.20) & 225(541)$^\mathrm{\star}$ & & &  \\
 & HCN1-0 &  0.53(0.09) & 228(50) & & &  \\
 & HCO+1-0 &  0.73(0.09) & 254(36) & & &  \\
 & HNC1-0 &  0.22(0.09) & 192(93) & & &  \\
 & HC3N10-9 &  $<$0.10 & - & & &  \\
& & & & & & \\ 
NGC4418 & H13CO+1-0 &  $<$0.32 & - & & &  \\
 & SiO2-1 &  0.32(0.14) & 150(33) & & &  \\
 & H13CN1-0 &  0.28(0.1) & 150(33) & & &  \\
 & C2H1-0 &  0.32(0.28) & 148(33)$^\mathrm{\star}$ & & &  \\
 & HCN1-0 &  2.15(0.15) & 188(33) & & &  \\
 & HCO+1-0 &  1.27(0.13) & 133(33) & & &  \\
 & HNC1-0 &  1.00(0.13) & 150(33) & & &  \\
 & HC3N10-9 &  0.80(0.13) & 187(33) & & &  \\
& & & & & & \\ 
NGC6090 & H13CO+1-0 &  $<$0.09 & - & & &  \\
 & SiO2-1 &  $<$0.09 & - & & &  \\
 & H13CN1-0 &  $<$0.09 & - & & &  \\
 & C2H1-0 &  0.22(0.06) & 103(30)$^\mathrm{\star}$ & & &  \\
 & HCN1-0 &  0.48(0.08) & 133(30) & & &  \\
 & HCO+1-0 &  0.80(0.08) & 128(30) & & &  \\
 & HNC1-0 &  0.12(0.06) & 91(30) & & &  \\
 & HC3N10-9 &  $<$0.09 & - & & &  \\
& & & & & & \\ 
NGC6240 & H13CO+1-0 &  $<$0.24 & - &  CH3OH &  $<$0.49 & - \\
 & SiO2-1 &  $<$0.24 & - &  HC3N12-11 &  $<$0.49 & - \\
 & H13CN1-0 &  $<$0.24 & - &  HC3N12-11v7e &  $<$0.49 & - \\
 & C2H1-0 &  0.94(0.16) & 674(126)$^\mathrm{\star}$ &  C18O1-0 &  1.00(0.18) & 400(70) \\
 & HCN1-0 &  2.79(0.15) & 493(33) &  13CO1-0 &  1.90(0.18) & 400(70) \\
 & HCO+1-0 &  4.54(0.16) & 541(24) &  CN1-0J=1/2 &  2.00(0.20) & 400(70) \\
 & HNC1-0 &  0.55(0.13) & 400(94) &  CN1-0J=3/2 &  3.70(0.20) & 400(70) \\
 & HC3N10-9 &  $<$0.24 & - &  CO1-0 &  54.70(0.50) & 414(70) \\
& & & & & & \\ 
NGC7469 & H13CO+1-0 &  $<$0.14 & - &  CH3OH &  0.16(0.14) & 55(407) \\
 & SiO2-1 &  $<$0.14 & - &  HC3N12-11 &  0.07(0.04) & 56(258) \\
 & H13CN1-0 &  $<$0.14 & - &  HC3N12-11v7e &  $<$0.22 & - \\
 & C2H1-0 &  1.50(0.40) & 503(53)$^\mathrm{\star}$ &  C18O1-0 &  0.33(0.07) & 209(41) \\
 & HCN1-0 &  1.97(0.10) & 248(12) &  13CO1-0 &  2.26(0.05) & 268(8) \\
 & HCO+1-0 &  2.24(0.10) & 346(11) &  CN1-0J=1/2 &  1.80(0.12) & 298(10) \\
 & HNC1-0 &  1.10(0.10) & 298(35) &  CN1-0J=3/2 &  2.80(0.14) & 278(14) \\
 & HC3N10-9 &  $<$0.14 & - &  CO1-0 &  47.00(0.02) & 274(1) \\
& & & & & & \\ 
NGC7771 & H13CO+1-0 &  $<$0.29 & - &  CH3OH &  0.44(0.14) & 293(87) \\
 & SiO2-1 &  0.23(0.36) & 275(400) &  HC3N12-11 &  $<$0.42 & - \\
 & H13CN1-0 & $<$0.29  & - &  HC3N12-11v7e &  $<$0.42 & - \\
 & C2H1-0 &  0.90(0.15) & 442(78)$^\mathrm{\star}$ &  C18O1-0 &  1.20(0.20) & 445(81) \\
 & HCN1-0 &  3.40(0.15) & 396(19) &  13CO1-0 &  5.60(0.20) & 459(16) \\
 & HCO+1-0 &  3.10(0.15) & 400(21) &  CN1-0J=1/2 &  1.12(0.24) & 400(30) \\
 & HNC1-0 &  2.00(0.16) & 469(44) &  CN1-0J=3/2 &  2.78(0.26) & 397(31) \\
 & HC3N10-9 &  0.52(0.12) & 400(100) &  CO1-0 &  77.60(0.24) & 422(2) \\
& & & & & & \\ 
\end{tabular}

\caption{Summary of the line properties of the observed galaxies ($continued$). See caption of Table \ref{tab:linprop}.}
\end{table*}
\begin{table*}
\centering
\begin{tabular}{l l c c | l c c} 
\hline 
\hline 
& & & & & & \\ 
 Galaxy & Transition & $\int T^{\star}_A dv$ & $\Delta_V$ & Transition & $\int T^{\star}_A dv$ & $\Delta_V$ \\ 
 & & K km s$^{-1}$ & km s$^{-1}$ & & K km s$^{-1}$ & s$^{-1}$ \\  
& & & & & & \\ 
\hline 
& & & & & & \\ 
& 88 GHz Band & & & 112 GHz Band & & \\ 
& & & & & & \\ 
NGC660 & H13CO+1-0 &  $<$0.35 & - &  CH3OH &  $<$0.35 & - \\
 & SiO2-1 &  $<$0.35 & - &  HC3N12-11 &  $<$0.35 & - \\
 & H13CN1-0 &  $<$0.35 & - &  HC3N12-11v7e &  $<$0.35 & - \\
 & C2H1-0 &  2.17(0.28) & 412(57)$^\mathrm{\star}$ &  C18O1-0 &  2.42(0.22) & 361(54) \\
 & HCN1-0 &  4.80(0.20) & 297(14) &  13CO1-0 &  6.70(0.18) & 299(54) \\
 & HCO+1-0 &  5.00(0.20) & 318(16) &  CN1-0J=1/2 &  4.10(0.18) & 334(54) \\
 & HNC1-0 &  2.50(0.20) & 323(33) &  CN1-0J=3/2 &  6.70(0.20) & 336(54) \\
 & HC3N10-9 &  $<$0.35 & - &  CO1-0 &  110.20(0.30) & 271(54) \\
& & & & & & \\ 
NGC3556 & H13CO+1-0 &  $<$0.15 & - &  CH3OH &  $<$0.46 & - \\
 & SiO2-1 &  $<$0.15 & - &  HC3N12-11 &  $<$0.46 & - \\
 & H13CN1-0 &  $<$0.15 & - &  HC3N12-11v7e &  $<$0.46 & - \\
 & C2H1-0 &  0.84(0.08) & 409(66)$^\mathrm{\star}$ &  C18O1-0 &  0.26(0.05) & 54(5) \\
 & HCN1-0 &  0.54(0.08) & 73(66) &  13CO1-0 &  3.36(0.06) & 63(5) \\
 & HCO+1-0 &  0.85(0.08) & 191(66) &  CN1-0J=1/2 &  0.40(0.10) & 103(5) \\
 & HNC1-0 &  0.16(0.08) & 81(66) &  CN1-0J=3/2 &  0.60(0.10) & 89(5) \\
 & HC3N10-9 &  $<$0.15 & - &  CO1-0 &  42.00(0.30) & 80(5) \\
& & & & & & \\ 
NGC1068 & H13CO+1-0 &  0.6(0.5) & 230(33) & & &  \\
 & SiO2-1 &  0.8(0.5) & 230(33) & & &  \\
 & H13CN1-0 &  1.26(0.5) & 230(33) & & &  \\
 & C2H1-0 &  8.10(1.80) & 471(33)$^\mathrm{\star}$ & & &  \\
 & HCN1-0 &  22.00(0.40) & 241(33) & & &  \\
 & HCO+1-0 &  14.70(0.40) & 232(33) & & &  \\
 & HNC1-0 &  8.90(0.40) & 238(33) & & &  \\
 & HC3N10-9 &  0.96(0.36) & 251(33) & & &  \\
& & & & & & \\ 
NGC7674 & H13CO+1-0 &  $<$0.18 & - &  CH3OH &  $<$0.51 & - \\
 & SiO2-1 &  $<$0.18 & - &  HC3N12-11 &  $<$0.51 & - \\
 & H13CN1-0 &  $<$0.18 & - &  HC3N12-11v7e &  $<$0.51 & - \\
 & C2H1-0 &  $<$0.18 & - &  C18O1-0 &  0.51(0.10) & 238(50) \\
 & HCN1-0 &  $<$0.18 & - &  13CO1-0 &  1.40(0.10) & 196(13) \\
 & HCO+1-0 &  $<$0.18 & - &  CN1-0J=1/2 &  0.51(0.14) & 238(62) \\
 & HNC1-0 &  $<$0.18 & - &  CN1-0J=3/2 &  1.10(0.20) & 317(67) \\
 & HC3N10-9 &  $<$0.18 & - &  CO1-0 &  20.25(0.20) & 200(16) \\
& & & & & & \\ 
UGC2866 & H13CO+1-0 &  $<$0.21 & - &  CH3OH &  $<$0.60 & - \\
 & SiO2-1 &  $<$0.21 & - &  HC3N12-11 &  $<$0.60 & - \\
 & H13CN1-0 &  $<$0.21 & - &  HC3N12-11v7e &  $<$0.60 & - \\
 & C2H1-0 &  1.30(0.15) & 397(50)$^\mathrm{\star}$ &  C18O1-0 &  0.60(0.14) & 283(56) \\
 & HCN1-0 &  1.47(0.12) & 254(25) &  13CO1-0 &  3.46(0.12) & 283(10) \\
 & HCO+1-0 &  2.15(0.12) & 273(15) &  CN1-0J=1/2 &  0.69(0.12) & 190(30) \\
 & HNC1-0 &  0.75(0.13) & 336(55) &  CN1-0J=3/2 &  1.98(0.16) & 264(22) \\
 & HC3N10-9 &  $<$0.21 & - &  CO1-0 &  71.54(0.28) & 266(1) \\
& & & & & & \\ 
UGC5101 & H13CO+1-0 &  $<$0.31 & - &  CH3OH &  $<$1.85 & - \\
 & SiO2-1 &  $<$0.31 & - &  HC3N12-11 &  $<$1.85 & - \\
 & H13CN1-0 &  0.37(0.17) & 456(130) &  HC3N12-11v7e &  $<$1.85 & - \\
 & C2H1-0 &  0.98(0.27) & 1043(130)$^\mathrm{\star}$ &  C18O1-0 &  $<$1.85 & - \\
 & HCN1-0 &  1.10(0.15) & 455(130) &  13CO1-0 &  $<$1.85 & - \\
 & HCO+1-0 &  0.40(0.16) & 500(130) &  CN1-0J=1/2 &  1.85(0.19) & 578(112) \\
 & HNC1-0 &  0.90(0.18) & 476(130) &  CN1-0J=3/2 &  1.85(0.19) & 578(112) \\
 & HC3N10-9 &  $<$0.31 & - &  CO1-0 &  16.20(0.30) & 536(112) \\
& & & & & & \\ 
NGC2273 & H13CO+1-0 &  $<$0.33 & - & & &  \\
 & SiO2-1 &  $<$0.33 & - & & &  \\
 & H13CN1-0 &  $<$0.33 & - & & &  \\
 & C2H1-0 &  $<$0.33 & - & & &  \\
 & HCN1-0 &  0.55(0.10) & 154(66) & & &  \\
 & HCO+1-0 &  0.33(0.10) & 154(66) & & &  \\
 & HNC1-0 &  0.33(0.10) & 154(66) & & &  \\
 & HC3N10-9 &  $<$0.33 & - & & &  \\
& & & & & & \\ 
\end{tabular}

\caption{Summary of the line properties of the observed galaxies ($continued$). See caption of Table \ref{tab:linprop}.}
\end{table*}
\begin{table*}
\centering
\begin{tabular}{l l c c | l c c} 
\hline 
\hline 
& & & & & & \\ 
 Galaxy & Transition & $\int T^{\star}_A dv$ & $\Delta_V$ & Transition & $\int T^{\star}_A dv$ & $\Delta_V$ \\ 
 & & K km s$^{-1}$ & km s$^{-1}$ & & K km s$^{-1}$ & s$^{-1}$ \\  
& & & & & & \\ 
\hline 
& & & & & & \\ 
& 88 GHz Band & & & 112 GHz Band & & \\ 
& & & & & & \\ 
NGC1377 & - &  $<$0.3 & 70 &   &  &  \\
& & & & & & \\ 
IRAS15250 & - &  $<$2.7 & 170 &  &   &  \\
& & & & & & \\ 
\end{tabular}

\caption{Upper limits (3-$\sigma$) for non-detections, estimated from noise $rms$. The assumed line width was derived from CO 1-0 observations.}
\end{table*}


\end{appendix}

\end{document}